\documentclass[reprint, twocolumn, amsmath, amssymb, superscriptaddress]{revtex4-2}

\usepackage{graphicx}
\usepackage{dcolumn}
\usepackage{amsmath}

\usepackage[usenames,dvipsnames]{xcolor} 
\usepackage{hhline}
\usepackage{array}
\newcolumntype{P}[1]{>{\centering\arraybackslash}p{#1}}
\usepackage{textcomp}
\usepackage{comment}
\usepackage[normalem]{ulem}

\begin{document}

\title{Precise and extensive characterization of an optical resonator for cavity-based quantum networks}


\author{Dowon Lee}
\affiliation{Department of Electrical Engineering, Pohang University of Science and Technology (POSTECH), 37673 Pohang, Korea}
\author{Myunghun Kim}
\affiliation{Department of Electrical Engineering, Pohang University of Science and Technology (POSTECH), 37673 Pohang, Korea}
\author{Jungsoo Hong}
\affiliation{Department of Electrical Engineering, Pohang University of Science and Technology (POSTECH), 37673 Pohang, Korea}
\author{Taegyu Ha}
\affiliation{Department of Electrical Engineering, Pohang University of Science and Technology (POSTECH), 37673 Pohang, Korea}
\author{Junwoo Kim}
\affiliation{Department of Electrical Engineering, Pohang University of Science and Technology (POSTECH), 37673 Pohang, Korea}
\author{Sungsam Kang}
\affiliation{Center for Molecular Spectroscopy and Dynamics, Institute for Basic Science (IBS), 145 Anam-ro, Seongbuk-gu, Seoul, 02841, Korea}
\affiliation{Department of Physics, Korea University, 145 Anam-ro, Seongbuk-gu, Seoul 02841, Korea}
\author{Youngwoon Choi}
\affiliation{Department of Bioengineering, Korea University, Seoul 02841, Korea}
\affiliation{Interdisciplinary Program in Precision Public Health, Korea University, Seoul 02841, Korea}
\author{Kyungwon An}
\affiliation{Department of Physics and Astronomy \& Institute of Applied Physics, Seoul National University, 08826 Seoul, Korea}
\author{Moonjoo Lee}
\email{moonjoo.lee@postech.ac.kr}
\affiliation{Department of Electrical Engineering, Pohang University of Science and Technology (POSTECH), 37673 Pohang, Korea}

\date{\today}

\begin{abstract}
Cavity-based quantum node is a competitive platform for distributed quantum networks.
Here, we characterize a high-finesse Fabry-P\'erot optical resonator for coupling single or few atomic quantum registers.
Our cavity consists of two mirrors with different reflectivities: One has minimal optical loss, and the other high transmission loss where more than 90\% of the intracavity photons would be emitted. 
Cavity finesse, birefringent effects, and mechanical resonances are measured using the lasers at 780, 782, and 795 nm.
In order to obtain cavity geometric parameters, we drive the adjacent longitudinal or transverse modes with two lasers simultaneously, and measure those frequencies using a precision wavelength meter (WLM).
A major novelty of this method is that the parameters' uncertainties are solely determined by the resolution of the WLM, eliminating all of the temporal environment fluctuations. 
Our scheme makes it possible to quantify the atom-cavity coupling constant up to four significant figures, the most precise and accurate estimation so far, which would become a key ingredient for benchmarking a cavity-based quantum node.
Furthermore, the distortion of polarized photonic qubits would be minimized owing to the small birefringent splitting, below 4.9\% of the cavity linewidth.
Our system should operate in the intermediate atom-cavity coupling regime that would allow us to implement various quantum network protocols.
\end{abstract}

\maketitle

\section{Introduction}

Quantum network~\cite{Wehner2018} is essential to realize a secure communication~\cite{Bennett2014} and construct a large-scale quantum computer~\cite{Monroe2014}.
In this network architecture, while individual nodes host stationary qubits, various quantum states are generated, teleported, and transfered among the nodes using photons, through communication channels like optical fibers~\cite{Northup14}. 
Such quantum state distribution is based on the generation of entanglement within, and across the interconnected nodes~\cite{Wilk07b, Stute12}. 
In current technology, the entanglement of stationary qubits inside a node was well-accomplished~\cite{Monz11} and characterized via thorough benchmarking~\cite{Erhard2019, Wright2019}.
On the other hand, it still remains a challenge to generate more efficient, higher-fidelity entanglement between stationary qubits and photons, and between remote quantum memories mediated by photons~\cite{Ritter12, Hucul2015, Stephenson2020, Langenfeld2021}.

\begin{figure*} [!t]
	\centering\includegraphics[width=13.2cm]{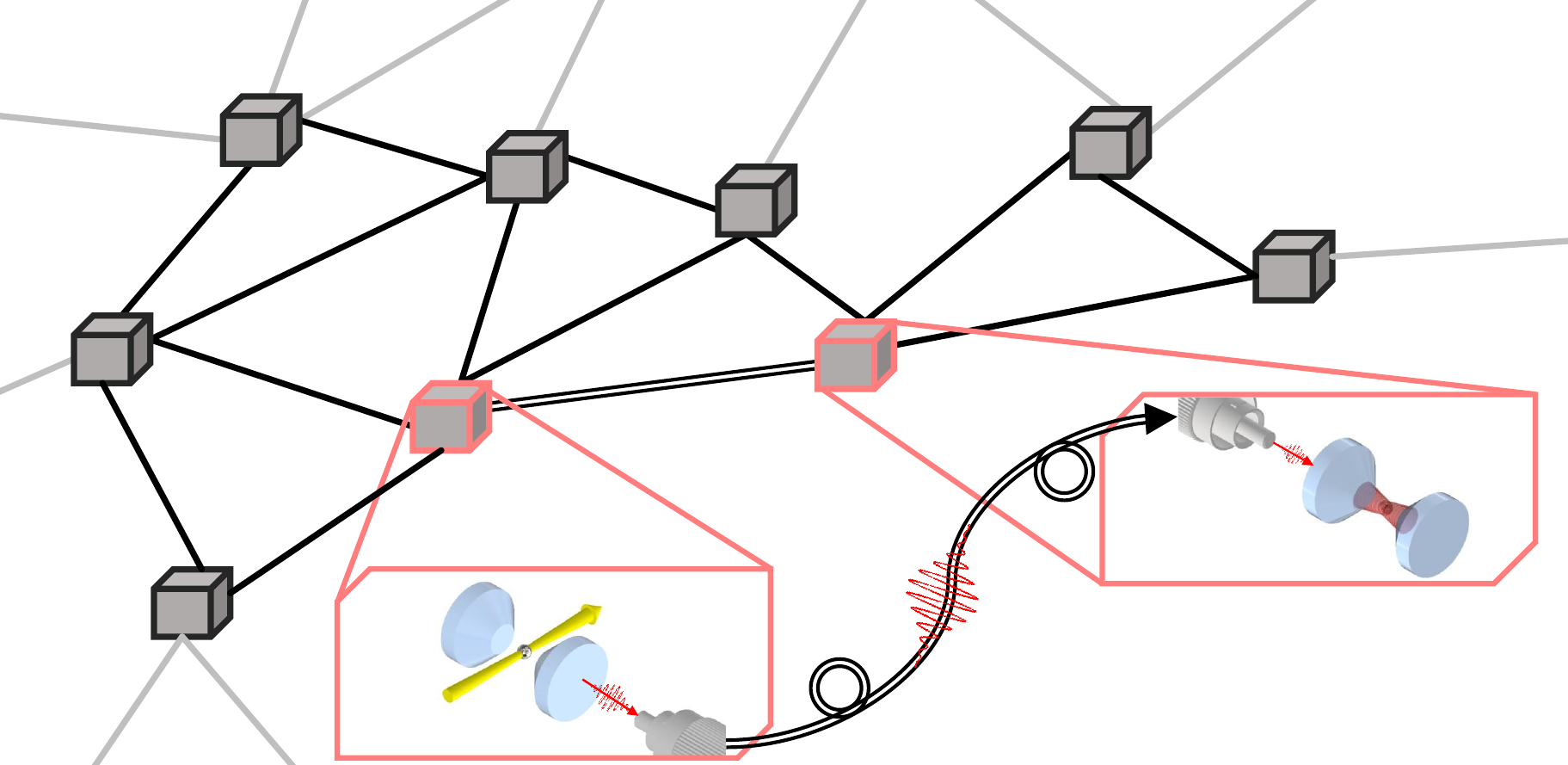} 
	\caption{
			Schematic of cavity-based quantum network. 
			Each quantum node consists of trapped atomic registers and optical cavity. 
			Stationary qubits are controlled by laser fields (yellow arrow); quantum states are distributed across the nodes mediated by the photons (red arrow) traveling through optical fibers.
	}
	\label{fig:network}
\end{figure*}

Among various experimental settings~\cite{Kimble08a}, cavity-based quantum node is a promising platform for realizing a distributed quantum network~\cite{Reiserer2015}.  
This system is comprised of single/few atoms or ions coupled to a high-finesse optical resonator. 
Here it is advantageous to utilize the capability of precise control of the atomic qubits~\cite{Harty14, Madjarov2020}, as well as their long coherence times~\cite{Koerber2018, Wang2021}.
Moreover, the enhanced cavity-vacuum fields~\cite{Lee2014} would increase the photon-collection efficiency and enable a deterministic energy exchange between atomic and photonic qubits~\cite{Kimble08a, Reiserer2015}.
Seminal experiments were reported with the cavity-based nodes, including the efficient generation of single photons~\cite{Muecke2013, Schupp2021}, deterministic link between remote nodes~\cite{Ritter12}, reversible quantum state transfer~\cite{Boozer07a, Stute13}, and gate operation using reflecting photons~\cite{Reiserer2014quantum, Hacker2016}.

The underlying mechanism of all these achievements consists of the coherent atom-photon interaction, which is quantified by the atom-cavity coupling constant $g_{0}$~\cite{Berman94, Kimble98}. 
This interaction not only governs the evolution of the internal states of the atomic or ion qubits~\cite{Wilk07b, Stute12a}, but also affects the external dynamics, such as atomic localization and cavity-induced cooling~\cite{Maunz04, Leibrandt09}.
It is therefore of fundamental importance to estimate $g_{0}$ for characterizing the cavity-based node in a quantitative way.
It becomes more crucial in a large-scale network, in which single photons are generated in several nodes simultaneously and entanglements are created in a concatenated way~\cite{Monroe2014, Kimble08a}.
In parallel, it is also needed to well-characterize other physical properties like the finesse and birefringence that influence the properties of the distributed quantum states.
Here, we report a precise, extensive, and self-contained characterization of a Fabry-P\'erot-type optical resonator for a cavity-based quantum network node. 
Two mirrors have different loss characteristics: One mirror with an ultralow optical loss, and the other high transmission loss through which most of the intracavity photons are outcoupled. 
The mirror substrate is shaped~\cite{Kim2012, Lee2014} such that laser beams are positioned closely to the cavity mirrors.
We determine the mirror spacing, mode waist, Rayleigh range, and radius of curvature (ROC) by measuring the frequency separations between adjacent TEM$_{00}$ modes and TEM$_{00}$--TEM$_{10}$ modes, where both frequencies are about 1~THz.
It is noteworthy that we make use of two lasers for exciting the modes and measure the wavelengths of the lasers simultaneously---which enable a precise quantification of all geometric parameters.
The atom-cavity coupling constant $g_{0}$ is estimated with all these values, at the highest precision and accuracy to date.
We stress that even the field leakage to the mirror coating is taken into account for the quantification of $g_0$.
We also measure the finesse of the cavity for several input polarizations, showing that the birefringent splitting is below 4.9\% of the cavity linewidth.
From the transmission measurement of the output mirror, we anticipate that the maximum outcoupling efficiency of an intracavity photon would be 92(4)\% at 780~nm.
Note that this outcoupling efficiency is defined as the probability that one intracavity photon is transmitted to the mirror of high-transmission loss.
Finally, we find that the cavity assembly's mechanical resonances are of frequencies higher than 20~kHz, agreeing with the results of the numerical simulation.

\begin{figure*} [!t]
	\centering\includegraphics[width=6.3in]{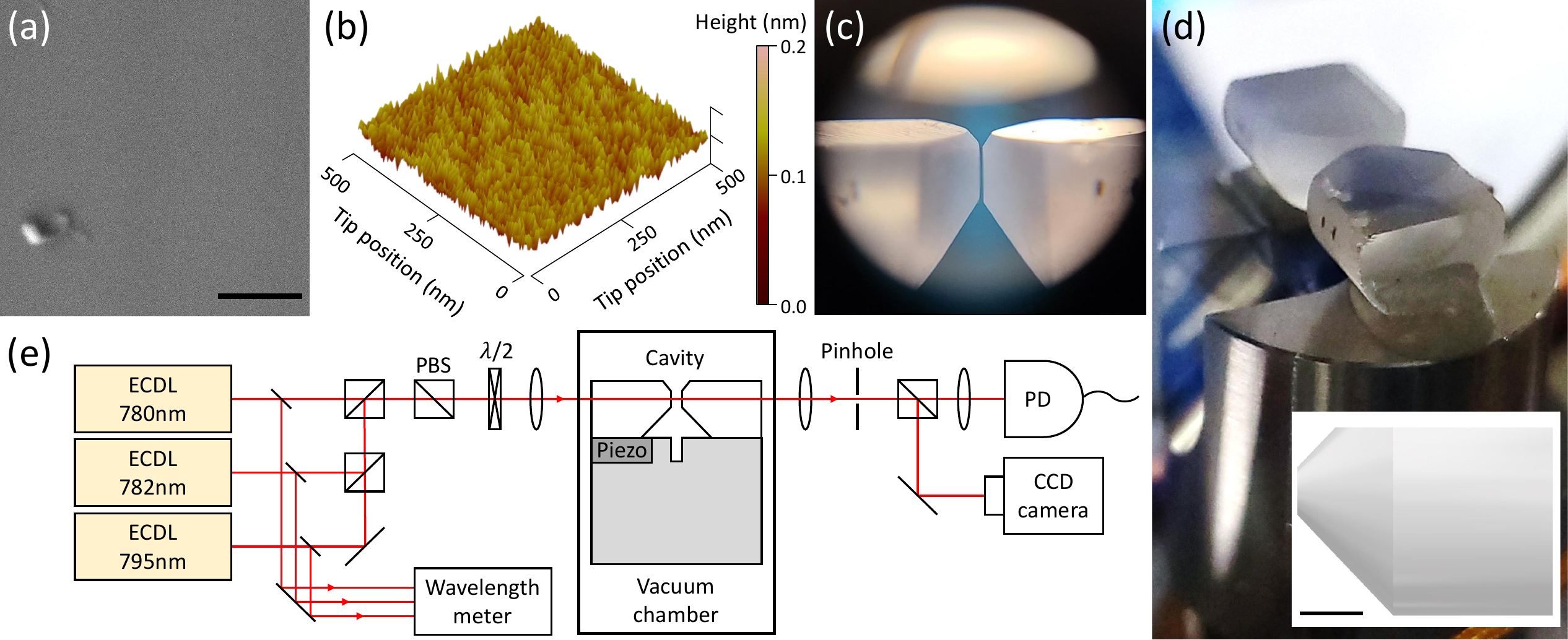} 
	\caption{
		(a) Scanning electron microscope image of mirror surface. 
		Nonuniform spot is dirt. 
		Scale bar denotes 1~$\mu$m.
		(b) Atomic force microscope image of mirror surface. 
		(c), (d) Photographs of assembled cavity under vacuum.
		Cavity length is $\sim$150~$\mu$m (see main text).		
		Inset in (d) shows cavity mirror design with a scale bar of 2~mm.
		(e) Experimental setup for cavity characterization. 
		External cavity diode laser (ECDL); Polarizing beam splitter (PBS); half-wave plate ($\lambda$/2); Photodiode (PD); Charge coupled device (CCD) camera.
	}
	\label{fig:design}
\end{figure*}

\section{Result}

\subsection{Experimental system}
Our cavity mirrors were coated in Research Electro-Optics (REO), Inc.: The coating of the ultralow loss mirror is centered at 791~nm, while the high-transmission mirror at 786~nm.
The reflectance band is  $\sim$10~nm for both mirrors.
The superpolished substrate of fused silica was fabricated to have a ROC $R=10$~cm  with a diameter of 7.75~mm and height of 8~mm.
The specification of the ultralow-loss mirror includes a transmission loss of 0.5--2~parts per million (ppm) and scattering loss below 5--10~ppm, while a transmission loss of the output mirror is 200~ppm.
As presented in Figs.~\ref{fig:design}(a) and (b), we begin with observing the mirror surface with a scanning electron microscope (SEM) and atomic force microscope (AFM). 
The AFM measurement yields a surface roughness of $\sim$0.4~\AA~over 500~nm$^2$.

The cylindrical mirror substrate is beveled to a `cone-like' shape until a mirror diameter of 1.5~mm remains.
We then grind further for flattening the upper area of the mirror substrate (Figs.~\ref{fig:design}(c) and (d)) such that a small atomic ensemble could be trapped on top of the cavity more closely. 
After aligning the two machined mirrors, we glue them to a cavity mount followed by room-temperature curing for several days. 
Next, the cavity assembly is installed in a vacuum chamber, and we proceed for baking at $\sim$100$^\circ$C for two weeks.
The chamber pressure is stabilized on the order of $10^{-10}$~mbar eventually.
After the experiments~\cite{Choi2010, Kang2011} with this setting, we moved the system from Seoul to Pohang in order to link with other quantum nodes, replaced the rubidium dispenser, baked the vacuum chamber again, and performed the characterization reported in this paper.
No degradation of the cavity finesse was found over the whole process.

Our experimental setup for the cavity characterization is depicted in Fig.~\ref{fig:design}(e).  
We employ three frequency-tunable external cavity diode lasers operating at 780, 782, and 795~nm~\cite{Kim2021} to drive the cavity: The two lasers at 780 and 782~nm are used for measuring the geometric parameters, and the laser at 795~nm for a finesse measurement at different wavelengths.
All laser frequencies are monitored in the WLM (HighFinesse, WS8-10 PCS8) with a frequency resolution of 10~MHz. 
The input polarization is defined by the polarizing beam splitter and $\lambda$/2 waveplate. 
The cavity transmission is measured with a photodiode, and the charge-coupled device camera is used for measuring the spatial profile of the mode. 
We control the cavity frequency by applying voltages to the shear-type piezoelectric transducer below the input mirror. 
The transducer displaces the mirror by one free spectral range (FSR) in $\sim$770~V. 
Note that the laser scattering of the driving field is spatially filtered with a pinhole between the cavity and the detectors.

\subsection{Cavity geometric parameters}
 
\subsubsection{Frequency measurements}
 
\begin{figure} [!t]
	\centering\includegraphics[width=3.3in]{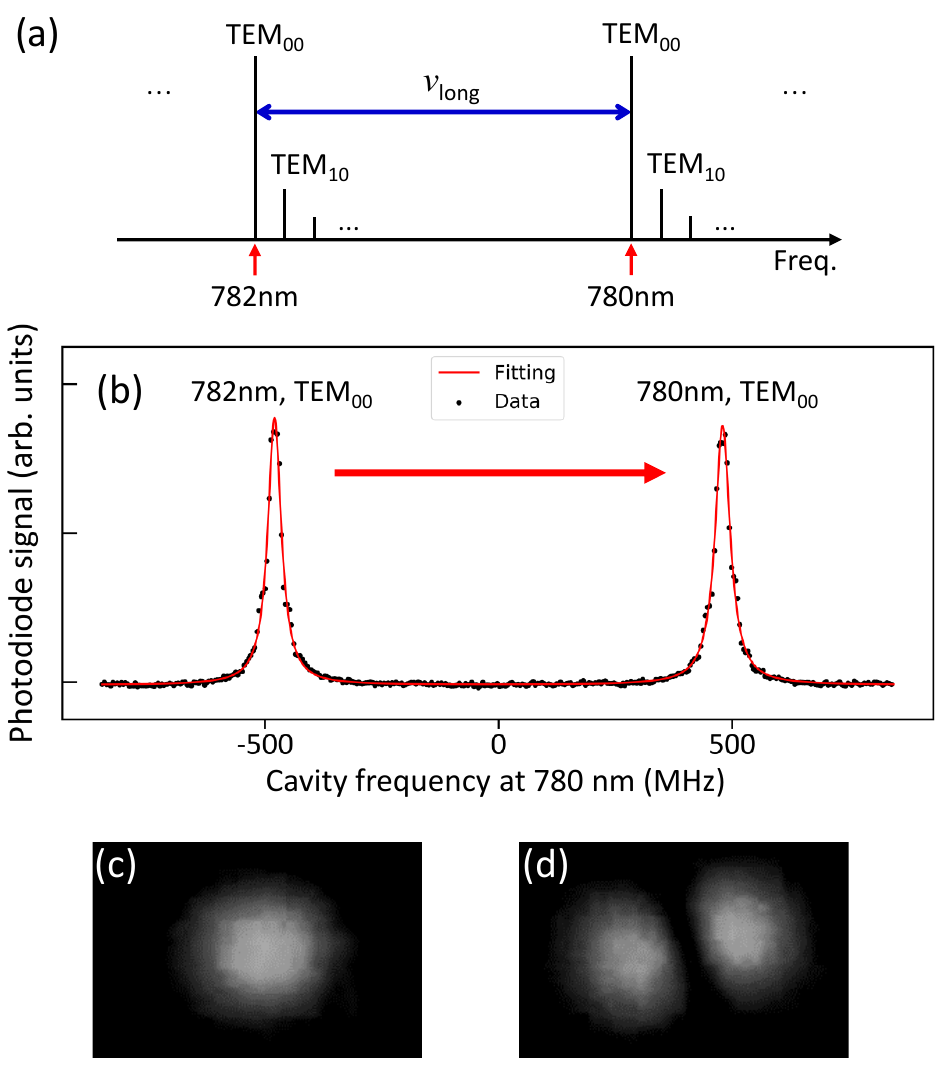} 
	\caption{
	(a) Longitudinal and transverse modes of the cavity.
	We drive the $n$th TEM$_{00}$ and TEM$_{10}$ modes with the laser at 782~nm, and the $(n+1)$th TEM$_{00}$ mode at 780~nm.
	(b) Photodiode signal as a function of the cavity frequency at 780~nm, while both lasers excite the adjacent longitudinal modes.  
	Actually we scan the cavity length by $\delta L$, and the horizontal axis of cavity frequency at 780~nm, $\delta\nu_{780}$, is the conversion from $\delta L$ to $\delta\nu_{780}$ via $\delta\nu_{780} \simeq -2c/ \left( (n+1)\lambda_{780}^2  \right) \times \delta L$, with resonant wavelength at 780~nm of $\lambda_{780}$.
	Left peak corresponds to the transmission of 782~nm laser and the right one 780~nm.
	We adjust frequencies of the two lasers such that both TEM$_{00}$ modes are excited simultaneously (red arrow). 
	CCD camera image of TEM$_{00}$ mode in (c) and TEM$_{10}$ mode in (d).	 
	}
	\label{fig:spectrum}
\end{figure}

We first measure the frequency difference between adjacent longitudinal modes. 
The cavity is driven with two lasers at 782 and 780~nm while scanning the cavity length.
We denote the frequency of each laser as $\nu_{782}$ and $\nu_{780}$, individually.
From the photograph shown in Fig.~\ref{fig:design}(c), we estimate the cavity length about 150~$\mu$m, and correspondingly $\nu_{\rm{fsr}}\simeq$1~THz. 
Since the frequency difference between the two lasers is $\sim$1~THz, we ensure that we are driving the adjacent longitudinal modes of the cavity.
As shown in Figs.~\ref{fig:spectrum}(a) and (b) we drive the $n$th and $(n+1)$th longitudinal modes (frequency of each mode $\nu_{n,00}$ and $\nu_{(n+1),00}$), where $n$ is the longitudinal mode index. 
Adjusting the frequencies of both lasers, we excite the two TEM$_{00}$ modes in such a way that $\nu_{n,00} = \nu_{782}$ and $\nu_{(n+1),00} = \nu_{780}$ at the same time.
Here the measured laser frequencies at the WLM are 383.23957(1)~THz (782.25863(2)~nm) and 384.22777(1)~THz (780.24672(2)~nm), resulting in a frequency difference of $\nu_{\rm{long}}=0.98820(1)$~THz. 
The measurement error is dominated by the WLM's frequency resolution. 

Next, we measure the frequency separation $\nu_{\rm{trans}}$ between a longitudinal TEM$_{00}$ and transverse TEM$_{10}$ mode. 
The procedure is similar with the FSR measurement: Driving the $(n+1)$th TEM$_{10}$ mode at a frequency of $\nu_{(n+1),10}$ using the 780~nm laser, we also excite $n$th TEM$_{00}$ mode with the 782~nm laser. 
Here we tune the laser frequencies so that $\nu_{(n+1),10}=\nu_{780}$ and $\nu_{n,00}=\nu_{782}$ simultaneously.
The WLM measurement gives a frequency separation of the two lasers of 1.00547(1)~THz. 
Subtracting $\nu_{\rm{long}}$ from this value, we obtain a frequency separation $\nu_{\rm{trans}}$ of 17.27(1)~GHz.
Note that the frequencies of TEM$_{10}$ and TEM$_{01}$ modes are degenerate in our cavity, with an upper bound of 0.9~MHz determined by the fitting error of the transmission peak.

\subsubsection{Extraction of cavity parameters}

Based on the aforementioned measurements, we extract the cavity geometric parameters.
We begin with the frequency of the $n$th longitudinal mode

\begin{equation}
	\nu_{n, 00} = \nu_{\rm{fsr}} \left( n - \frac{\phi_{\rm{coat}}(\nu_{n,00})}{2\pi} + \frac{\Delta\zeta}{\pi} \right)
\end{equation}

\noindent
where $\nu_{\rm{fsr}}=c/2L$,  $c$ is the speed of light, $L$ is the distance between the facets of two mirrors, $\phi_{\rm{coat}}(\nu_{n,00})$ is the phase shift occurred in the mirror coating, and $\Delta\zeta$ is the Gouy phase originated from the nature of Gaussian beam~\cite{Feng2001, Kim2012, Siegman86}.
Here we define $L_{\rm{eff}}$ as an effective cavity length, which includes the effect of field leakage to the mirror coating and the ROC of the mirrors, i.e., 

\begin{equation}
	\nu_{\rm{long}} = \nu_{(n+1), 00} -\nu_{n, 00} =  \frac{c}{2L_{\rm{eff}}}.
	\label{eq:Leff}
\end{equation}

\noindent
From Eq.~(\ref{eq:Leff}) and the measurement of $\nu_{\rm{long}}$, we obtain $L_{\rm{eff}} = 151.686(2)$~$\mu$m. 
Now we rule out the influence of $\phi_{\rm{coat}}(\nu_{n,00})$ for extracting other parameters. 
As studied in Ref.~\cite{Hood01}, electromagnetic field in the mirror coating induces an additional phase shift of $\phi_{\rm{coat}}(\nu_{n,00})$ that varies with the wavelength. 
We numerically calculate this phase difference $\Delta\phi_{\rm{coat}} = \phi_{\rm{coat}}(\nu_{(n+1), 00}) -\phi_{\rm{coat}}(\nu_{n, 00})$ of 782 and 780~nm via the transfer matrix method~\cite{Yeh2005, Hood01, Garcia2020}, where we obtain $\Delta\phi_{\rm{coat}} = 29.3132(4)$~mrad (see Appendix).
It is then straightforward to obtain the relation $\nu_{\rm{long}} = \nu_{\rm{fsr}}(1-\Delta\phi_{\rm{coat}}/2\pi)$, and therefore we find $\nu_{\rm{fsr}}$ = 0.99283(1)~THz and $L = 150.978(2)$~$\mu$m.

We continue with the estimation of $R$. 
Employing the formula $\nu_{\rm{trans}}=\Delta\zeta \cdot \nu_{\rm{fsr}} /\pi$~\cite{Siegman86}, a phase shift of $\Delta\zeta = 54.65(3)$~mrad is estimated, resulting in $R=10.124(1)$~cm through the relation $\Delta\zeta = 2\tan^{-1}{(\sqrt{L/(2R-L)})}$.
The obtained $R$, the effective ROC that includes the contribution of both mirrors, shows a good agreement with the REO Inc.'s specification: Quoted from the mirror vendor, the uncertainty of the ROC is about 1\%.
We then proceed to obtain the mode waist $w_0= (L(2R-L)\lambda^2 / 4\pi^2)^{1/4} = 26.198(8)$~$\mu$m at 780~nm, and 26.232(8)~$\mu$m at 782~nm.
The Rayleigh range $z_{0}=\pi w_{0}^2/\lambda$~\cite{Siegman86} is given by $z_{0}=2.763(2)$~mm.

Lastly, we discuss the difference between the actual mirror spacing $L$ and effective cavity length $L_{\rm{eff}}$.
While all other geometric parameters above are determined by $L$, the mode volume should be governed by $L_{\rm{eff}}$ that includes the effect of the electromagnetic field in the coating layers. 
We obtain the cavity mode volume V$_{c} = \pi w_0^2 L_{\rm{eff}}/4 =81.77(5)$~picoliters (pL) at 780~nm and 81.98(5)~pL at 782~nm ($L\ll z_{0}$).

\subsection{Finesse}

We continue with the measurement of the cavity finesse. 
The cavity transmission of a TEM$_{00}$ mode is fit to a Lorentzian function, which yields a full width at half maximum (FWHM) of 37.1(9)~MHz through averaging the results of 76 experiments.
The error is attributed to the shot-to-shot fluctuations of the mechanical vibrations and acoustic noises. 
The obtained FWHM corresponds to $2\kappa$ where $\kappa$ is the cavity-field decay rate.
We do not find the difference of the linewidth for the wavelengths at 780 and 782~nm.
Using the relation $F=\pi c/(2\kappa L_{\rm{eff}})$, we obtain a cavity finesse of $2.66(6)\times10^{4}$ at both wavelengths.
We also measure the cavity finesse at 795~nm, yielding a slightly higher finesse of $2.84(7)\times10^{4}$.
We would attribute this finesse difference to the wavelength-dependent scattering loss (possibly some dirts on the mirror facet), or diffraction loss due to the slight cavity misalighment.

\subsection{Birefringence}

Birefringence properties of our cavity are then investigated. 
The birefringent effect originates from mechanical, thermal, or inherent stress in the mirror substrate or coating, resulting in different optical path lengths for distinct light polarizations~\cite{Huang2008, Barrett2019, Morales2019}.
This effect would be detrimental for encoding photonic qubit information into the polarization, because it distorts the polarization qubit when coupled to the cavity mode. 
We carry out two types of measurements in order to check such properties.
First, we would measure the polarization shift (or distortion) of the output field with respect to that of an input field for identifying the fast or slow axis of the cavity. 
We drive a TEM$_{00}$ mode of the cavity with a horizontal polarization H and analyzed the output polarization. We find that the intensity of vertical polarization component is below the noise level ($\sim$10$^{-3}$ of the signal) of our detection system; a similar measurement is done with several linear input polarizations, giving rise to negligible detection of orthogonal component in the output field.
We expected that when the cavity is driven along one of the polarization eigenmodes, the phase retardation, correspondingly the polarization shift/distortion, would be minimized; if the cavity is driven with a polarization that excites both eigenmodes identically, the polarization change would be maximized.
Our measurement could not tell the cavity fast or slow axis.

Second, we drive a TEM$_{00}$ mode with three input polarizations of H, V, and (H+V)$/\sqrt{2}$ at 780 and 795~nm, and analyze the FWHM of the cavity transmission peaks. 
This measurement is to check whether the cavity field of different polarizations would exhibit separate resonance frequencies. 
If the birefringent splitting $\nu_{\rm{brf}}$ were to be measurable, two distinct peaks should appear at least for one of the three input polarizations.
We do not find any distortion of the transmission, and all the cavity transmission peaks are well fitted to single Lorentzian functions of the same FWHM with the value in Sec.~2.3.
The obtained results show that our cavity's birefringence effect is below the level of our measurement precision.
More quantitatively, we can set the upper bound of $\nu_{\rm{brf}}$ as the error bar of the FWHM measurement, 0.9~MHz, which corresponds to 4.9\% of $\kappa$ at both wavelengths.
In a future study, we plan to measure the birefringence more carefully, e.g. following the approach of Refs.~\cite{Dupre2015, Fleisher2016} that combines the cavity ringdown measurement with a polarization analysis.

We then compare the birefringence of our resonator to that of other similar settings.
The splitting $\nu_{\rm{brf}}$ was similar with the cavity linewidth~\cite{Birnbaum05a}, and about 10$\%$ of cavity linewidth in neutral single atom-cavity experiments~\cite{Hacker2016}.
In ion-cavity experiments, the effect was not noticeable~\cite{Stute12} and around 10\% of cavity linewidth~\cite{Krutyanskiy2019}.  
Interestingly, the birefringence effect was exploited for observing time oscillations of single photon polarization~\cite{Barrett2019} and for realizing two-mode Dicke model~\cite{Morales2019}.


\subsection{Outcoupling efficiency}
\begin{figure} [htbp]
	\centering\includegraphics[width=3.3in]{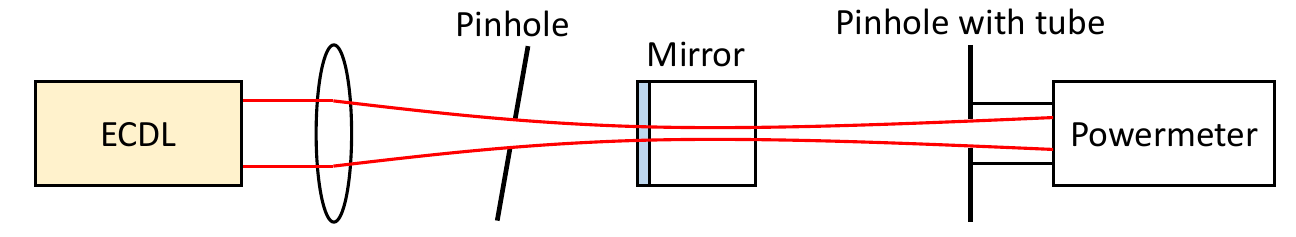} 
	\caption{
	Experimental setup for transmission measurement of output mirror. 
	}
	\label{fig:transmission}
\end{figure}

In order to attain a maximum photon collection efficiency, we develop the cavity so that most of the cavity photons are emitted through the mirror with high transmission loss.
This outcoupling efficiency (see our definition in Sec.~1) is estimated by combining the results of finesse measurement with a separate calibration: We measure the transmission of 8 mirrors in the same coating run as the output mirror, using the setting shown in Fig.~\ref{fig:transmission}. 
The anodized pinhole in front of the mirror is tilted for preventing the back-reflection/scattering, at the surface of the pinhole, from propagating to the powermeter.
In the experiment, we tilt the pinhole by $\sim$10$^\circ$, check that such influence is negligible, and measure the transmittance.
The averaged transmittance $T$ is 218(9)~ppm at 780~nm and 220(4)~ppm at 795~nm. 
The uncertainties are dominated by the statistical error that corresponds to the standard deviation of the transmittances of all measured mirrors. 
The contribution of the systematic error, the linearity of the powermeter, is 0.8\% of the statistical error in the 780-nm mirror measurement and 2\% of the statistical one in the 795-nm measurement.
Given the total optical loss of our cavity $L_{\rm{tot}}=2\pi/F$ is 236(6)~ppm at 780~nm and 221(6)~ppm at 795~nm, we acquire outcoupling efficiencies of $T/L_{\rm{tot}}$ of $0.92(4)$ and $0.99^{+1}_{-3}$ at each wavelength.
When generating single photons with this system, single atoms in the cavity are driven by a laser from an orthogonal direction with respect to the cavity axis. 
The obtained $T/L_{\rm{tot}}$ corresponds to the upper bound of the single-photon generation efficiency: This efficiency is reduced from the outcoupling efficiency by the probability of atomic emission to the free space, imperfect excitation of the atoms, optical loss from the cavity to the photodetector, and limited quantum efficiency of the detector.

\subsection{Mechanical mode}

\begin{figure*} [!t]
	\centering\includegraphics[width=6.3in]{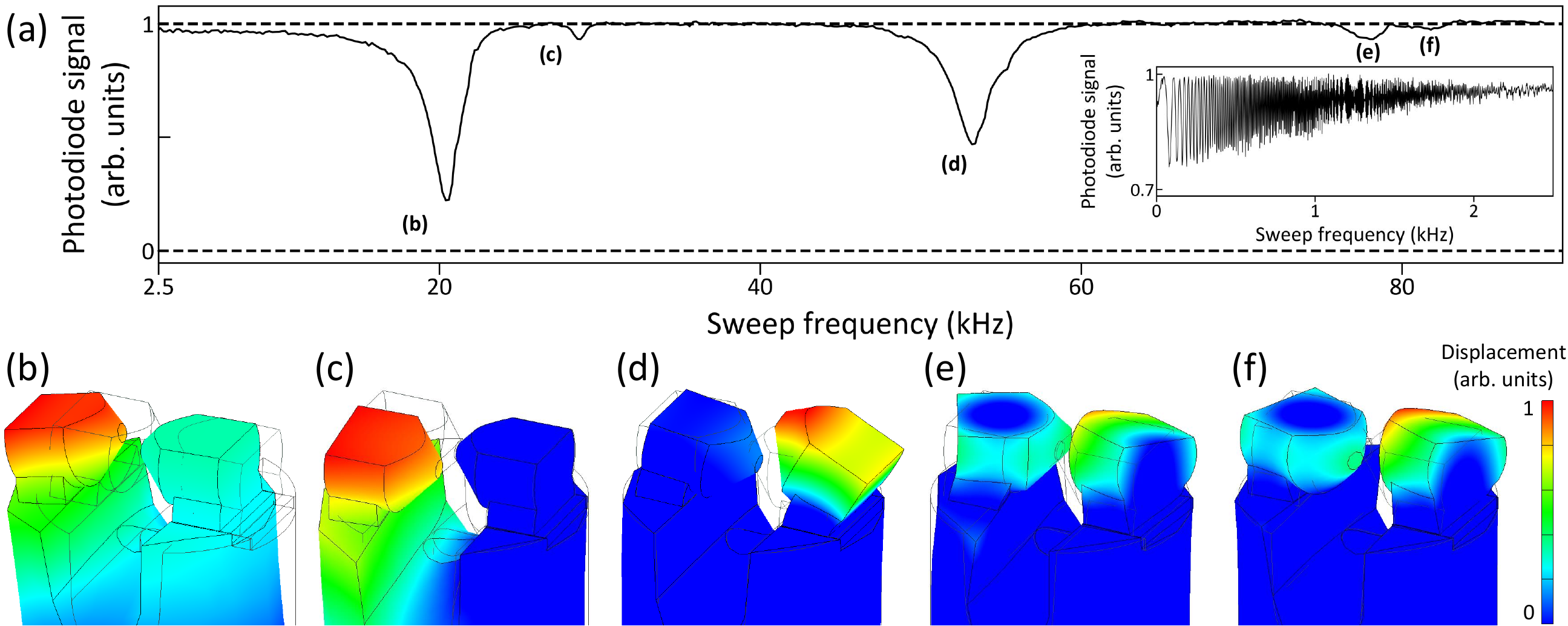} 
	\caption{
		(a) Cavity transmission (solid line) as a function of the sweeping frequency to piezoelectric transducer. 
		Two dashed lines indicate lower and upper bounds of normalization of the transmission.
		Slightly low transmission at $<10$~kHz is attributed to slow drift of cavity frequency ($\sim$3~MHz/s).
		Inset: Low frequency response of the cavity. 
	 	(b)--(f) Numerical modal analysis of mechanical resonances.	
	}
	\label{fig:mech_reso}
\end{figure*}

Our last characterization consists of measuring mechanical resonances of the cavity assembly. 
For stabilizing the cavity frequency, it is crucial to have high mechanical frequencies since it would determine the maximum feedback frequencies~\cite{Briles2010, Janitz2017, Lee2019a}: The phase shift and nonlinear amplitude response near the resonances cause instabilities in the frequency stabilization~\cite{Gallego2016}.
For measuring such resonances, we drive the cavity with the 780~nm laser on resonance to a TEM$_{00}$ mode. 
We then apply a frequency-sweeping (chirping) voltage $V(t)$ to the shear-type piezoelectric transducer from $f_{\rm{i}}=0$ to $f_{\rm{f}}=90$~kHz in $T=0.5$~s, i.e., $V(t)=V_{0}\sin \left(    2\pi      (  (f_{\rm{f}}-f_{\rm{i}})t /T  )            t      \right)$ where $V_{0}=10$~mV. 
The cavity transmission decreases when the sweeping frequency is close to the mechanical resonance: The oscillation amplitude of the mirror increases, and thus the cavity frequency becomes non-resonant from the laser frequency.
As presented in Fig.~\ref{fig:mech_reso}(a), two dominant responses appear at sweep frequencies of 21 and 53~kHz with other subsidiary transmission dips. 
Those two pronounced decreases of the transmission (Figs.~\ref{fig:mech_reso}(b) and (d)) occur when the motion of the cavity mirrors is along the cavity axis, where the displacement changes the cavity length in the first order of the amplitude.
Other smaller transmission drops appear when the mirrors' oscillation direction is mostly orthogonal to the cavity axis. 

All mechanical resonances could be explained by the finite-element, numerical modal analysis using Autodesk{\textsuperscript{\tiny\textregistered}} Inventor Professional 2021 (Figs.~\ref{fig:mech_reso}(b)-(f)).
From the minimum of transmission dips, we identify the mechanical resonances at 21, 29, 53, 78, and 82~kHz, with a good agreement of the simulation results of 21, 28, 54, 78, and 80~kHz.
Note that the low frequency response (inset of Fig.~\ref{fig:mech_reso}(a)) does not exhibit any feature of mechanical resonances.
This measurement shows that our cavity frequency stabilization can be done with a bandwidth $<20$~kHz.

\section{Discussion}

The maximum atom-cavity coupling constant is then estimated in our cavity.
We consider a single $^{87}$Rb atom located at an antinode in the waist of a TEM$_{00}$ mode of the cavity.
When the D$_2$ transition 5$^{2}$S$_{1/2}$$ |F=2\rangle$~$\leftrightarrow$~5$^{2}$P$_{3/2}$$|F'=3\rangle$ at 780~nm is coupled to the cavity frequency, we obtain
\begin{equation*}
	g_{0}/2\pi = \left( \frac{3}{2\sqrt{2} \pi^2} \right)^{1/2}  \sqrt{  \frac{ c\lambda}{\sqrt{R L_{\rm{eff}}^{3}  }  }   \left( \frac{\gamma}{\pi} \right)     } = 16.02(1)~\rm{MHz}  
\end{equation*}
where the atomic decay rate $2\gamma= 2\pi \cdot 6.065(9)$~MHz~\cite{Steck2010}. 
Our atom-cavity system is in the imtermediate coupling regime with a critical photon number of $\gamma^2/(2g_{0}^{2})=0.01793(6)$ and cooperativity parameter of $g_0^{2}/ (2\kappa \gamma)=2.28(6)$~\cite{Berman94}.

We compare our method to other techniques and point out the major novelty of our characterization in two aspects.
First, our method eliminates most of the technical noises for determining the cavity geometric parameters, so that the attained uncertainty is only determined by the resolution of the WLM. 
When characterizing a resonator for cavity QED experiments, it is common practice that cavity or laser frequencies are scanned using a piezoelectric actuator to measure the frequency between cavity modes.
In typical laboratory environment, the frequency jitter over the scanning duration is governed by electrical, mechanical, or acoustic noises in uncontrollable ways.
Since we measure two laser frequencies simultaneously, temporal environment fluctuations would be ruled out completely and thus, our characterization makes it possible to estimate $g_0$ up to four significant figures --- this is the most precise and accurate estimation so far, to the best of our knowledge.
We employ the WLM with a frequency resolution of 10~MHz, which allows us to reveal the cavity length and Gouy phase at uncertainties as low as $2$~nm and 15~$\mu$rad, respectively.
More precision and accuracy could be obtained with the currently available WLM of a higher frequency resolution~\cite{Ghadimi2020}. 
Second, our method is useful for short cavities used in cavity QED experiments.
Since the FSR is on the order of hundreds of GHz or THz for such cavities~\cite{Ritter12, Brandstaetter2013, Ann2019, Macha2020, Takahashi2020}, it is not straightforward to extract the cavity parameters via scanning laser/cavity frequencies, or electro-optically modulating sidebands. 

We remark the difference between our work and other two experiments~\cite{Hunger2010, Neuzner2016thesis}.
In Ref.~\cite{Hunger2010}, adjacent longitudinal modes of a fiber cavity were driven with two lasers for measuring a FSR. 
However, the frequency resolution of their six-digit Burleigh WLM was $\sim$490~MHz at 780~nm, which is larger than that of ours by about a factor of 50.
They measured the cavity length with an uncertainty below 500~nm, while our uncertainty is on the order of nanometers. 
Next, in Ref.~\cite{Neuzner2016thesis}, a high-resolution WLM was employed like our experiment, however, while they drove two longitudinal modes of the cavity (like Ref.~\cite{Hunger2010}), transverse modes were not driven, which would make it not possible to estimate the Gouy phase and ROC of the cavity mirrors. 
Therefore they took the value of the ROC of the mirror vendor to estimate $g_{0}$.
In contrast, we extracted all geometric parameters via driving both longitudinal and transverse modes.

Our last note includes the reason that we made use of the laser at 782~nm.
The extraction of cavity geometric parameters should have been possible with any pair of lasers, e.g. 780~nm and 795~nm lasers if two longitudinal modes could be driven simultaneously. 
In our configuration, however, it was not the case owing to the limited scan range of both lasers, $\sim$130 GHz. 
We therefore exploit the laser at 782~nm, which can drive two adjacent TEM$_{00}$ modes with the 780~nm laser.

\section{Conclusion}

In conclusion, we have developed and extensively characterized a high-finesse Fabry-P\'erot optical cavity for coupling single neutral atoms.  
We measure the cavity length and finesse with three different lasers operating at 780, 782, and 795~nm.
We anticipate that more than 90\% of intracavity photons would be emitted through the highly transmissive mirror with negligible polarization distortions. 
No significant mechanical modes are found below 20~kHz, which would be helpful for stabilizing the cavity frequency.
Our method could be exploited for characterizing not only the cavities for atom- or ion-based quantum networks~\cite{Ritter12, Brandstaetter2013, Ong2020, Takahashi2020}, but also other Fabry-P\'erot cavities in which fundamental quantum optics~\cite{Lee2019}, precision measurement~\cite{Norcia2016}, and many-body physics are investigated~\cite{Baumann10, Leonard2017, Hosseini2017}.

\section{Acknowledgments}
We thank Jae-Yoon Sim for experimental support and Josef Schupp for helpful discussions.
K.~An was supported by the National Research Foundation (Grant No.~2020R1A2C3009299).
This work has been supported by National Research Foundation (Grant No.~2019R1A5A102705513 and 2020R1l1A2066622), BK21 FOUR program, Samsung Science and Technology Foundation (SSTF-BA2101-07 and SRFC-TC2103-01), and Samsung Electronics Co., Ltd (IO201211-08121-01).

\section{Data availability} Data underlying the results presented in this paper are available in Ref.~\cite{dowon_lee_2022_6324425}

\bibliographystyle{apsrev4-2}
\bibliography{bibliography}

\begin{thebibliography}{66}%
\makeatletter
\providecommand \@ifxundefined [1]{%
 \@ifx{#1\undefined}
}%
\providecommand \@ifnum [1]{%
 \ifnum #1\expandafter \@firstoftwo
 \else \expandafter \@secondoftwo
 \fi
}%
\providecommand \@ifx [1]{%
 \ifx #1\expandafter \@firstoftwo
 \else \expandafter \@secondoftwo
 \fi
}%
\providecommand \natexlab [1]{#1}%
\providecommand \enquote  [1]{``#1''}%
\providecommand \bibnamefont  [1]{#1}%
\providecommand \bibfnamefont [1]{#1}%
\providecommand \citenamefont [1]{#1}%
\providecommand \href@noop [0]{\@secondoftwo}%
\providecommand \href [0]{\begingroup \@sanitize@url \@href}%
\providecommand \@href[1]{\@@startlink{#1}\@@href}%
\providecommand \@@href[1]{\endgroup#1\@@endlink}%
\providecommand \@sanitize@url [0]{\catcode `\\12\catcode `\$12\catcode
  `\&12\catcode `\#12\catcode `\^12\catcode `\_12\catcode `\%12\relax}%
\providecommand \@@startlink[1]{}%
\providecommand \@@endlink[0]{}%
\providecommand \url  [0]{\begingroup\@sanitize@url \@url }%
\providecommand \@url [1]{\endgroup\@href {#1}{\urlprefix }}%
\providecommand \urlprefix  [0]{URL }%
\providecommand \Eprint [0]{\href }%
\providecommand \doibase [0]{https://doi.org/}%
\providecommand \selectlanguage [0]{\@gobble}%
\providecommand \bibinfo  [0]{\@secondoftwo}%
\providecommand \bibfield  [0]{\@secondoftwo}%
\providecommand \translation [1]{[#1]}%
\providecommand \BibitemOpen [0]{}%
\providecommand \bibitemStop [0]{}%
\providecommand \bibitemNoStop [0]{.\EOS\space}%
\providecommand \EOS [0]{\spacefactor3000\relax}%
\providecommand \BibitemShut  [1]{\csname bibitem#1\endcsname}%
\let\auto@bib@innerbib\@empty
\bibitem [{\citenamefont {Wehner}\ \emph {et~al.}(2018)\citenamefont {Wehner},
  \citenamefont {Elkouss},\ and\ \citenamefont {Hanson}}]{Wehner2018}%
  \BibitemOpen
  \bibfield  {author} {\bibinfo {author} {\bibfnamefont {S.}~\bibnamefont
  {Wehner}}, \bibinfo {author} {\bibfnamefont {D.}~\bibnamefont {Elkouss}},\
  and\ \bibinfo {author} {\bibfnamefont {R.}~\bibnamefont {Hanson}},\
  }\href@noop {} {\bibfield  {journal} {\bibinfo  {journal} {Science}\ }\textbf
  {\bibinfo {volume} {362}} (\bibinfo {year} {2018})}\BibitemShut {NoStop}%
\bibitem [{\citenamefont {Bennett}\ and\ \citenamefont
  {Brassard}(2014)}]{Bennett2014}%
  \BibitemOpen
  \bibfield  {author} {\bibinfo {author} {\bibfnamefont {C.~H.}\ \bibnamefont
  {Bennett}}\ and\ \bibinfo {author} {\bibfnamefont {G.}~\bibnamefont
  {Brassard}},\ }\href
  {https://doi.org/https://doi.org/10.1016/j.tcs.2014.05.025} {\bibfield
  {journal} {\bibinfo  {journal} {Theor. Comput. Sci.}\ }\textbf {\bibinfo
  {volume} {560}},\ \bibinfo {pages} {7 } (\bibinfo {year} {2014})}\BibitemShut
  {NoStop}%
\bibitem [{\citenamefont {Monroe}\ \emph {et~al.}(2014)\citenamefont {Monroe},
  \citenamefont {Raussendorf}, \citenamefont {Ruthven}, \citenamefont {Brown},
  \citenamefont {Maunz}, \citenamefont {Duan},\ and\ \citenamefont
  {Kim}}]{Monroe2014}%
  \BibitemOpen
  \bibfield  {author} {\bibinfo {author} {\bibfnamefont {C.}~\bibnamefont
  {Monroe}}, \bibinfo {author} {\bibfnamefont {R.}~\bibnamefont {Raussendorf}},
  \bibinfo {author} {\bibfnamefont {A.}~\bibnamefont {Ruthven}}, \bibinfo
  {author} {\bibfnamefont {K.~R.}\ \bibnamefont {Brown}}, \bibinfo {author}
  {\bibfnamefont {P.}~\bibnamefont {Maunz}}, \bibinfo {author} {\bibfnamefont
  {L.-M.}\ \bibnamefont {Duan}},\ and\ \bibinfo {author} {\bibfnamefont
  {J.}~\bibnamefont {Kim}},\ }\href
  {https://doi.org/10.1103/PhysRevA.89.022317} {\bibfield  {journal} {\bibinfo
  {journal} {Phys. Rev. A}\ }\textbf {\bibinfo {volume} {89}},\ \bibinfo
  {pages} {022317} (\bibinfo {year} {2014})}\BibitemShut {NoStop}%
\bibitem [{\citenamefont {Northup}\ and\ \citenamefont
  {Blatt}(2014)}]{Northup14}%
  \BibitemOpen
  \bibfield  {author} {\bibinfo {author} {\bibfnamefont {T.~E.}\ \bibnamefont
  {Northup}}\ and\ \bibinfo {author} {\bibfnamefont {R.}~\bibnamefont
  {Blatt}},\ }\href {https://doi.org/10.1038/nphoton.2014.53} {\bibfield
  {journal} {\bibinfo  {journal} {Nat. Photonics}\ }\textbf {\bibinfo {volume}
  {8}},\ \bibinfo {pages} {356} (\bibinfo {year} {2014})}\BibitemShut {NoStop}%
\bibitem [{\citenamefont {Wilk}\ \emph {et~al.}(2007)\citenamefont {Wilk},
  \citenamefont {Webster}, \citenamefont {Kuhn},\ and\ \citenamefont
  {Rempe}}]{Wilk07b}%
  \BibitemOpen
  \bibfield  {author} {\bibinfo {author} {\bibfnamefont {T.}~\bibnamefont
  {Wilk}}, \bibinfo {author} {\bibfnamefont {S.~C.}\ \bibnamefont {Webster}},
  \bibinfo {author} {\bibfnamefont {A.}~\bibnamefont {Kuhn}},\ and\ \bibinfo
  {author} {\bibfnamefont {G.}~\bibnamefont {Rempe}},\ }\href@noop {}
  {\bibfield  {journal} {\bibinfo  {journal} {Science}\ }\textbf {\bibinfo
  {volume} {317}},\ \bibinfo {pages} {488} (\bibinfo {year}
  {2007})}\BibitemShut {NoStop}%
\bibitem [{\citenamefont {Stute}\ \emph
  {et~al.}(2012{\natexlab{a}})\citenamefont {Stute}, \citenamefont {Casabone},
  \citenamefont {Schindler}, \citenamefont {Monz}, \citenamefont {Schmidt},
  \citenamefont {Brandst{\"a}tter}, \citenamefont {Northup},\ and\
  \citenamefont {Blatt}}]{Stute12}%
  \BibitemOpen
  \bibfield  {author} {\bibinfo {author} {\bibfnamefont {A.}~\bibnamefont
  {Stute}}, \bibinfo {author} {\bibfnamefont {B.}~\bibnamefont {Casabone}},
  \bibinfo {author} {\bibfnamefont {P.}~\bibnamefont {Schindler}}, \bibinfo
  {author} {\bibfnamefont {T.}~\bibnamefont {Monz}}, \bibinfo {author}
  {\bibfnamefont {P.~O.}\ \bibnamefont {Schmidt}}, \bibinfo {author}
  {\bibfnamefont {B.}~\bibnamefont {Brandst{\"a}tter}}, \bibinfo {author}
  {\bibfnamefont {T.~E.}\ \bibnamefont {Northup}},\ and\ \bibinfo {author}
  {\bibfnamefont {R.}~\bibnamefont {Blatt}},\ }\href
  {https://doi.org/10.1038/nature11120} {\bibfield  {journal} {\bibinfo
  {journal} {Nature}\ }\textbf {\bibinfo {volume} {485}},\ \bibinfo {pages}
  {482} (\bibinfo {year} {2012}{\natexlab{a}})}\BibitemShut {NoStop}%
\bibitem [{\citenamefont {Monz}\ \emph {et~al.}(2011)\citenamefont {Monz},
  \citenamefont {Schindler}, \citenamefont {Barreiro}, \citenamefont {Chwalla},
  \citenamefont {Nigg}, \citenamefont {Coish}, \citenamefont {Harlander},
  \citenamefont {H\"ansel}, \citenamefont {Hennrich},\ and\ \citenamefont
  {Blatt}}]{Monz11}%
  \BibitemOpen
  \bibfield  {author} {\bibinfo {author} {\bibfnamefont {T.}~\bibnamefont
  {Monz}}, \bibinfo {author} {\bibfnamefont {P.}~\bibnamefont {Schindler}},
  \bibinfo {author} {\bibfnamefont {J.~T.}\ \bibnamefont {Barreiro}}, \bibinfo
  {author} {\bibfnamefont {M.}~\bibnamefont {Chwalla}}, \bibinfo {author}
  {\bibfnamefont {D.}~\bibnamefont {Nigg}}, \bibinfo {author} {\bibfnamefont
  {W.~A.}\ \bibnamefont {Coish}}, \bibinfo {author} {\bibfnamefont
  {M.}~\bibnamefont {Harlander}}, \bibinfo {author} {\bibfnamefont
  {W.}~\bibnamefont {H\"ansel}}, \bibinfo {author} {\bibfnamefont
  {M.}~\bibnamefont {Hennrich}},\ and\ \bibinfo {author} {\bibfnamefont
  {R.}~\bibnamefont {Blatt}},\ }\href
  {https://doi.org/10.1103/PhysRevLett.106.130506} {\bibfield  {journal}
  {\bibinfo  {journal} {Phys. Rev. Lett.}\ }\textbf {\bibinfo {volume} {106}},\
  \bibinfo {pages} {130506} (\bibinfo {year} {2011})}\BibitemShut {NoStop}%
\bibitem [{\citenamefont {Erhard}\ \emph {et~al.}(2019)\citenamefont {Erhard},
  \citenamefont {Wallman}, \citenamefont {Postler}, \citenamefont {Meth},
  \citenamefont {Stricker}, \citenamefont {Martinez}, \citenamefont
  {Schindler}, \citenamefont {Monz}, \citenamefont {Emerson},\ and\
  \citenamefont {Blatt}}]{Erhard2019}%
  \BibitemOpen
  \bibfield  {author} {\bibinfo {author} {\bibfnamefont {A.}~\bibnamefont
  {Erhard}}, \bibinfo {author} {\bibfnamefont {J.~J.}\ \bibnamefont {Wallman}},
  \bibinfo {author} {\bibfnamefont {L.}~\bibnamefont {Postler}}, \bibinfo
  {author} {\bibfnamefont {M.}~\bibnamefont {Meth}}, \bibinfo {author}
  {\bibfnamefont {R.}~\bibnamefont {Stricker}}, \bibinfo {author}
  {\bibfnamefont {E.~A.}\ \bibnamefont {Martinez}}, \bibinfo {author}
  {\bibfnamefont {P.}~\bibnamefont {Schindler}}, \bibinfo {author}
  {\bibfnamefont {T.}~\bibnamefont {Monz}}, \bibinfo {author} {\bibfnamefont
  {J.}~\bibnamefont {Emerson}},\ and\ \bibinfo {author} {\bibfnamefont
  {R.}~\bibnamefont {Blatt}},\ }\href
  {https://doi.org/10.1038/s41467-019-13068-7} {\bibfield  {journal} {\bibinfo
  {journal} {Nat. Commun.}\ }\textbf {\bibinfo {volume} {10}},\ \bibinfo
  {pages} {5347} (\bibinfo {year} {2019})}\BibitemShut {NoStop}%
\bibitem [{\citenamefont {Wright}\ \emph {et~al.}(2019)\citenamefont {Wright},
  \citenamefont {Beck}, \citenamefont {Debnath}, \citenamefont {Amini},
  \citenamefont {Nam}, \citenamefont {Grzesiak}, \citenamefont {Chen},
  \citenamefont {Pisenti}, \citenamefont {Chmielewski}, \citenamefont
  {Collins}, \citenamefont {Hudek}, \citenamefont {Mizrahi}, \citenamefont
  {Wong-Campos}, \citenamefont {Allen}, \citenamefont {Apisdorf}, \citenamefont
  {Solomon}, \citenamefont {Williams}, \citenamefont {Ducore}, \citenamefont
  {Blinov}, \citenamefont {Kreikemeier}, \citenamefont {Chaplin}, \citenamefont
  {Keesan}, \citenamefont {Monroe},\ and\ \citenamefont {Kim}}]{Wright2019}%
  \BibitemOpen
  \bibfield  {author} {\bibinfo {author} {\bibfnamefont {K.}~\bibnamefont
  {Wright}}, \bibinfo {author} {\bibfnamefont {K.~M.}\ \bibnamefont {Beck}},
  \bibinfo {author} {\bibfnamefont {S.}~\bibnamefont {Debnath}}, \bibinfo
  {author} {\bibfnamefont {J.~M.}\ \bibnamefont {Amini}}, \bibinfo {author}
  {\bibfnamefont {Y.}~\bibnamefont {Nam}}, \bibinfo {author} {\bibfnamefont
  {N.}~\bibnamefont {Grzesiak}}, \bibinfo {author} {\bibfnamefont {J.-S.}\
  \bibnamefont {Chen}}, \bibinfo {author} {\bibfnamefont {N.~C.}\ \bibnamefont
  {Pisenti}}, \bibinfo {author} {\bibfnamefont {M.}~\bibnamefont
  {Chmielewski}}, \bibinfo {author} {\bibfnamefont {C.}~\bibnamefont
  {Collins}}, \bibinfo {author} {\bibfnamefont {K.~M.}\ \bibnamefont {Hudek}},
  \bibinfo {author} {\bibfnamefont {J.}~\bibnamefont {Mizrahi}}, \bibinfo
  {author} {\bibfnamefont {J.~D.}\ \bibnamefont {Wong-Campos}}, \bibinfo
  {author} {\bibfnamefont {S.}~\bibnamefont {Allen}}, \bibinfo {author}
  {\bibfnamefont {J.}~\bibnamefont {Apisdorf}}, \bibinfo {author}
  {\bibfnamefont {P.}~\bibnamefont {Solomon}}, \bibinfo {author} {\bibfnamefont
  {M.}~\bibnamefont {Williams}}, \bibinfo {author} {\bibfnamefont {A.~M.}\
  \bibnamefont {Ducore}}, \bibinfo {author} {\bibfnamefont {A.}~\bibnamefont
  {Blinov}}, \bibinfo {author} {\bibfnamefont {S.~M.}\ \bibnamefont
  {Kreikemeier}}, \bibinfo {author} {\bibfnamefont {V.}~\bibnamefont
  {Chaplin}}, \bibinfo {author} {\bibfnamefont {M.}~\bibnamefont {Keesan}},
  \bibinfo {author} {\bibfnamefont {C.}~\bibnamefont {Monroe}},\ and\ \bibinfo
  {author} {\bibfnamefont {J.}~\bibnamefont {Kim}},\ }\href
  {https://doi.org/10.1038/s41467-019-13534-2} {\bibfield  {journal} {\bibinfo
  {journal} {Nat. Commun.}\ }\textbf {\bibinfo {volume} {10}},\ \bibinfo
  {pages} {5464} (\bibinfo {year} {2019})}\BibitemShut {NoStop}%
\bibitem [{\citenamefont {Ritter}\ \emph {et~al.}(2012)\citenamefont {Ritter},
  \citenamefont {N\"olleke}, \citenamefont {Hahn}, \citenamefont {Reiserer},
  \citenamefont {Neuzner}, \citenamefont {Uphoff}, \citenamefont {M\"ucke},
  \citenamefont {Figueroa}, \citenamefont {Bochmann},\ and\ \citenamefont
  {Rempe}}]{Ritter12}%
  \BibitemOpen
  \bibfield  {author} {\bibinfo {author} {\bibfnamefont {S.}~\bibnamefont
  {Ritter}}, \bibinfo {author} {\bibfnamefont {C.}~\bibnamefont {N\"olleke}},
  \bibinfo {author} {\bibfnamefont {C.}~\bibnamefont {Hahn}}, \bibinfo {author}
  {\bibfnamefont {A.}~\bibnamefont {Reiserer}}, \bibinfo {author}
  {\bibfnamefont {A.}~\bibnamefont {Neuzner}}, \bibinfo {author} {\bibfnamefont
  {M.}~\bibnamefont {Uphoff}}, \bibinfo {author} {\bibfnamefont
  {M.}~\bibnamefont {M\"ucke}}, \bibinfo {author} {\bibfnamefont
  {E.}~\bibnamefont {Figueroa}}, \bibinfo {author} {\bibfnamefont
  {J.}~\bibnamefont {Bochmann}},\ and\ \bibinfo {author} {\bibfnamefont
  {G.}~\bibnamefont {Rempe}},\ }\href {https://doi.org/10.1038/nature11023}
  {\bibfield  {journal} {\bibinfo  {journal} {Nature}\ }\textbf {\bibinfo
  {volume} {484}},\ \bibinfo {pages} {195} (\bibinfo {year}
  {2012})}\BibitemShut {NoStop}%
\bibitem [{\citenamefont {Hucul}\ \emph {et~al.}(2015)\citenamefont {Hucul},
  \citenamefont {Inlek}, \citenamefont {Vittorini}, \citenamefont {Crocker},
  \citenamefont {Debnath}, \citenamefont {Clark},\ and\ \citenamefont
  {Monroe}}]{Hucul2015}%
  \BibitemOpen
  \bibfield  {author} {\bibinfo {author} {\bibfnamefont {D.}~\bibnamefont
  {Hucul}}, \bibinfo {author} {\bibfnamefont {I.~V.}\ \bibnamefont {Inlek}},
  \bibinfo {author} {\bibfnamefont {G.}~\bibnamefont {Vittorini}}, \bibinfo
  {author} {\bibfnamefont {C.}~\bibnamefont {Crocker}}, \bibinfo {author}
  {\bibfnamefont {S.}~\bibnamefont {Debnath}}, \bibinfo {author} {\bibfnamefont
  {S.~M.}\ \bibnamefont {Clark}},\ and\ \bibinfo {author} {\bibfnamefont
  {C.}~\bibnamefont {Monroe}},\ }\href {https://doi.org/10.1038/nphys3150}
  {\bibfield  {journal} {\bibinfo  {journal} {Nat. Phys.}\ }\textbf {\bibinfo
  {volume} {11}},\ \bibinfo {pages} {37} (\bibinfo {year} {2015})}\BibitemShut
  {NoStop}%
\bibitem [{\citenamefont {Stephenson}\ \emph {et~al.}(2020)\citenamefont
  {Stephenson}, \citenamefont {Nadlinger}, \citenamefont {Nichol},
  \citenamefont {An}, \citenamefont {Drmota}, \citenamefont {Ballance},
  \citenamefont {Thirumalai}, \citenamefont {Goodwin}, \citenamefont {Lucas},\
  and\ \citenamefont {Ballance}}]{Stephenson2020}%
  \BibitemOpen
  \bibfield  {author} {\bibinfo {author} {\bibfnamefont {L.~J.}\ \bibnamefont
  {Stephenson}}, \bibinfo {author} {\bibfnamefont {D.~P.}\ \bibnamefont
  {Nadlinger}}, \bibinfo {author} {\bibfnamefont {B.~C.}\ \bibnamefont
  {Nichol}}, \bibinfo {author} {\bibfnamefont {S.}~\bibnamefont {An}}, \bibinfo
  {author} {\bibfnamefont {P.}~\bibnamefont {Drmota}}, \bibinfo {author}
  {\bibfnamefont {T.~G.}\ \bibnamefont {Ballance}}, \bibinfo {author}
  {\bibfnamefont {K.}~\bibnamefont {Thirumalai}}, \bibinfo {author}
  {\bibfnamefont {J.~F.}\ \bibnamefont {Goodwin}}, \bibinfo {author}
  {\bibfnamefont {D.~M.}\ \bibnamefont {Lucas}},\ and\ \bibinfo {author}
  {\bibfnamefont {C.~J.}\ \bibnamefont {Ballance}},\ }\href
  {https://doi.org/10.1103/PhysRevLett.124.110501} {\bibfield  {journal}
  {\bibinfo  {journal} {Phys. Rev. Lett.}\ }\textbf {\bibinfo {volume} {124}},\
  \bibinfo {pages} {110501} (\bibinfo {year} {2020})}\BibitemShut {NoStop}%
\bibitem [{\citenamefont {Langenfeld}\ \emph {et~al.}(2021)\citenamefont
  {Langenfeld}, \citenamefont {Welte}, \citenamefont {Hartung}, \citenamefont
  {Daiss}, \citenamefont {Thomas}, \citenamefont {Morin}, \citenamefont
  {Distante},\ and\ \citenamefont {Rempe}}]{Langenfeld2021}%
  \BibitemOpen
  \bibfield  {author} {\bibinfo {author} {\bibfnamefont {S.}~\bibnamefont
  {Langenfeld}}, \bibinfo {author} {\bibfnamefont {S.}~\bibnamefont {Welte}},
  \bibinfo {author} {\bibfnamefont {L.}~\bibnamefont {Hartung}}, \bibinfo
  {author} {\bibfnamefont {S.}~\bibnamefont {Daiss}}, \bibinfo {author}
  {\bibfnamefont {P.}~\bibnamefont {Thomas}}, \bibinfo {author} {\bibfnamefont
  {O.}~\bibnamefont {Morin}}, \bibinfo {author} {\bibfnamefont
  {E.}~\bibnamefont {Distante}},\ and\ \bibinfo {author} {\bibfnamefont
  {G.}~\bibnamefont {Rempe}},\ }\href
  {https://doi.org/10.1103/PhysRevLett.126.130502} {\bibfield  {journal}
  {\bibinfo  {journal} {Phys. Rev. Lett.}\ }\textbf {\bibinfo {volume} {126}},\
  \bibinfo {pages} {130502} (\bibinfo {year} {2021})}\BibitemShut {NoStop}%
\bibitem [{\citenamefont {Kimble}(2008)}]{Kimble08a}%
  \BibitemOpen
  \bibfield  {author} {\bibinfo {author} {\bibfnamefont {H.~J.}\ \bibnamefont
  {Kimble}},\ }\href@noop {} {\bibfield  {journal} {\bibinfo  {journal}
  {Nature}\ }\textbf {\bibinfo {volume} {453}},\ \bibinfo {pages} {1023}
  (\bibinfo {year} {2008})}\BibitemShut {NoStop}%
\bibitem [{\citenamefont {Reiserer}\ and\ \citenamefont
  {Rempe}(2015)}]{Reiserer2015}%
  \BibitemOpen
  \bibfield  {author} {\bibinfo {author} {\bibfnamefont {A.}~\bibnamefont
  {Reiserer}}\ and\ \bibinfo {author} {\bibfnamefont {G.}~\bibnamefont
  {Rempe}},\ }\href {https://doi.org/10.1103/RevModPhys.87.1379} {\bibfield
  {journal} {\bibinfo  {journal} {Rev. Mod. Phys.}\ }\textbf {\bibinfo {volume}
  {87}},\ \bibinfo {pages} {1379} (\bibinfo {year} {2015})}\BibitemShut
  {NoStop}%
\bibitem [{\citenamefont {Harty}\ \emph {et~al.}(2014)\citenamefont {Harty},
  \citenamefont {Allcock}, \citenamefont {Ballance}, \citenamefont {Guidoni},
  \citenamefont {Janacek}, \citenamefont {Linke}, \citenamefont {Stacey},\ and\
  \citenamefont {Lucas}}]{Harty14}%
  \BibitemOpen
  \bibfield  {author} {\bibinfo {author} {\bibfnamefont {T.~P.}\ \bibnamefont
  {Harty}}, \bibinfo {author} {\bibfnamefont {D.~T.~C.}\ \bibnamefont
  {Allcock}}, \bibinfo {author} {\bibfnamefont {C.~J.}\ \bibnamefont
  {Ballance}}, \bibinfo {author} {\bibfnamefont {L.}~\bibnamefont {Guidoni}},
  \bibinfo {author} {\bibfnamefont {H.~A.}\ \bibnamefont {Janacek}}, \bibinfo
  {author} {\bibfnamefont {N.~M.}\ \bibnamefont {Linke}}, \bibinfo {author}
  {\bibfnamefont {D.~N.}\ \bibnamefont {Stacey}},\ and\ \bibinfo {author}
  {\bibfnamefont {D.~M.}\ \bibnamefont {Lucas}},\ }\href
  {https://doi.org/10.1103/PhysRevLett.113.220501} {\bibfield  {journal}
  {\bibinfo  {journal} {Phys. Rev. Lett.}\ }\textbf {\bibinfo {volume} {113}},\
  \bibinfo {pages} {220501} (\bibinfo {year} {2014})}\BibitemShut {NoStop}%
\bibitem [{\citenamefont {Madjarov}\ \emph {et~al.}(2020)\citenamefont
  {Madjarov}, \citenamefont {Covey}, \citenamefont {Shaw}, \citenamefont
  {Choi}, \citenamefont {Kale}, \citenamefont {Cooper}, \citenamefont
  {Pichler}, \citenamefont {Schkolnik}, \citenamefont {Williams},\ and\
  \citenamefont {Endres}}]{Madjarov2020}%
  \BibitemOpen
  \bibfield  {author} {\bibinfo {author} {\bibfnamefont {I.~S.}\ \bibnamefont
  {Madjarov}}, \bibinfo {author} {\bibfnamefont {J.~P.}\ \bibnamefont {Covey}},
  \bibinfo {author} {\bibfnamefont {A.~L.}\ \bibnamefont {Shaw}}, \bibinfo
  {author} {\bibfnamefont {J.}~\bibnamefont {Choi}}, \bibinfo {author}
  {\bibfnamefont {A.}~\bibnamefont {Kale}}, \bibinfo {author} {\bibfnamefont
  {A.}~\bibnamefont {Cooper}}, \bibinfo {author} {\bibfnamefont
  {H.}~\bibnamefont {Pichler}}, \bibinfo {author} {\bibfnamefont
  {V.}~\bibnamefont {Schkolnik}}, \bibinfo {author} {\bibfnamefont {J.~R.}\
  \bibnamefont {Williams}},\ and\ \bibinfo {author} {\bibfnamefont
  {M.}~\bibnamefont {Endres}},\ }\href
  {https://doi.org/10.1038/s41567-020-0903-z} {\bibfield  {journal} {\bibinfo
  {journal} {Nat. Phys.}\ }\textbf {\bibinfo {volume} {16}},\ \bibinfo {pages}
  {857} (\bibinfo {year} {2020})}\BibitemShut {NoStop}%
\bibitem [{\citenamefont {K{\"o}rber}\ \emph {et~al.}(2018)\citenamefont
  {K{\"o}rber}, \citenamefont {Morin}, \citenamefont {Langenfeld},
  \citenamefont {Neuzner}, \citenamefont {Ritter},\ and\ \citenamefont
  {Rempe}}]{Koerber2018}%
  \BibitemOpen
  \bibfield  {author} {\bibinfo {author} {\bibfnamefont {M.}~\bibnamefont
  {K{\"o}rber}}, \bibinfo {author} {\bibfnamefont {O.}~\bibnamefont {Morin}},
  \bibinfo {author} {\bibfnamefont {S.}~\bibnamefont {Langenfeld}}, \bibinfo
  {author} {\bibfnamefont {A.}~\bibnamefont {Neuzner}}, \bibinfo {author}
  {\bibfnamefont {S.}~\bibnamefont {Ritter}},\ and\ \bibinfo {author}
  {\bibfnamefont {G.}~\bibnamefont {Rempe}},\ }\href
  {https://doi.org/10.1038/s41566-017-0050-y} {\bibfield  {journal} {\bibinfo
  {journal} {Nat. Photonics}\ }\textbf {\bibinfo {volume} {12}},\ \bibinfo
  {pages} {18} (\bibinfo {year} {2018})}\BibitemShut {NoStop}%
\bibitem [{\citenamefont {Wang}\ \emph {et~al.}(2021)\citenamefont {Wang},
  \citenamefont {Luan}, \citenamefont {Qiao}, \citenamefont {Um}, \citenamefont
  {Zhang}, \citenamefont {Wang}, \citenamefont {Yuan}, \citenamefont {Gu},
  \citenamefont {Zhang},\ and\ \citenamefont {Kim}}]{Wang2021}%
  \BibitemOpen
  \bibfield  {author} {\bibinfo {author} {\bibfnamefont {P.}~\bibnamefont
  {Wang}}, \bibinfo {author} {\bibfnamefont {C.-Y.}\ \bibnamefont {Luan}},
  \bibinfo {author} {\bibfnamefont {M.}~\bibnamefont {Qiao}}, \bibinfo {author}
  {\bibfnamefont {M.}~\bibnamefont {Um}}, \bibinfo {author} {\bibfnamefont
  {J.}~\bibnamefont {Zhang}}, \bibinfo {author} {\bibfnamefont
  {Y.}~\bibnamefont {Wang}}, \bibinfo {author} {\bibfnamefont {X.}~\bibnamefont
  {Yuan}}, \bibinfo {author} {\bibfnamefont {M.}~\bibnamefont {Gu}}, \bibinfo
  {author} {\bibfnamefont {J.}~\bibnamefont {Zhang}},\ and\ \bibinfo {author}
  {\bibfnamefont {K.}~\bibnamefont {Kim}},\ }\href
  {https://doi.org/10.1038/s41467-020-20330-w} {\bibfield  {journal} {\bibinfo
  {journal} {Nat. Commun.}\ }\textbf {\bibinfo {volume} {12}},\ \bibinfo
  {pages} {233} (\bibinfo {year} {2021})}\BibitemShut {NoStop}%
\bibitem [{\citenamefont {Lee}\ \emph {et~al.}(2014)\citenamefont {Lee},
  \citenamefont {Kim}, \citenamefont {Seo}, \citenamefont {Hong}, \citenamefont
  {Song}, \citenamefont {Dasari},\ and\ \citenamefont {An}}]{Lee2014}%
  \BibitemOpen
  \bibfield  {author} {\bibinfo {author} {\bibfnamefont {M.}~\bibnamefont
  {Lee}}, \bibinfo {author} {\bibfnamefont {J.}~\bibnamefont {Kim}}, \bibinfo
  {author} {\bibfnamefont {W.}~\bibnamefont {Seo}}, \bibinfo {author}
  {\bibfnamefont {H.-G.}\ \bibnamefont {Hong}}, \bibinfo {author}
  {\bibfnamefont {Y.}~\bibnamefont {Song}}, \bibinfo {author} {\bibfnamefont
  {R.~R.}\ \bibnamefont {Dasari}},\ and\ \bibinfo {author} {\bibfnamefont
  {K.}~\bibnamefont {An}},\ }\href {https://doi.org/10.1038/ncomms4441}
  {\bibfield  {journal} {\bibinfo  {journal} {Nat. Commun.}\ }\textbf {\bibinfo
  {volume} {5}},\ \bibinfo {pages} {3441} (\bibinfo {year} {2014})}\BibitemShut
  {NoStop}%
\bibitem [{\citenamefont {M\"ucke}\ \emph {et~al.}(2013)\citenamefont
  {M\"ucke}, \citenamefont {Bochmann}, \citenamefont {Hahn}, \citenamefont
  {Neuzner}, \citenamefont {N\"olleke}, \citenamefont {Reiserer}, \citenamefont
  {Rempe},\ and\ \citenamefont {Ritter}}]{Muecke2013}%
  \BibitemOpen
  \bibfield  {author} {\bibinfo {author} {\bibfnamefont {M.}~\bibnamefont
  {M\"ucke}}, \bibinfo {author} {\bibfnamefont {J.}~\bibnamefont {Bochmann}},
  \bibinfo {author} {\bibfnamefont {C.}~\bibnamefont {Hahn}}, \bibinfo {author}
  {\bibfnamefont {A.}~\bibnamefont {Neuzner}}, \bibinfo {author} {\bibfnamefont
  {C.}~\bibnamefont {N\"olleke}}, \bibinfo {author} {\bibfnamefont
  {A.}~\bibnamefont {Reiserer}}, \bibinfo {author} {\bibfnamefont
  {G.}~\bibnamefont {Rempe}},\ and\ \bibinfo {author} {\bibfnamefont
  {S.}~\bibnamefont {Ritter}},\ }\href
  {https://doi.org/10.1103/PhysRevA.87.063805} {\bibfield  {journal} {\bibinfo
  {journal} {Phys. Rev. A}\ }\textbf {\bibinfo {volume} {87}},\ \bibinfo
  {pages} {063805} (\bibinfo {year} {2013})}\BibitemShut {NoStop}%
\bibitem [{\citenamefont {Schupp}\ \emph {et~al.}(2021)\citenamefont {Schupp},
  \citenamefont {Krcmarsky}, \citenamefont {Krutyanskiy}, \citenamefont
  {Meraner}, \citenamefont {Northup},\ and\ \citenamefont
  {Lanyon}}]{Schupp2021}%
  \BibitemOpen
  \bibfield  {author} {\bibinfo {author} {\bibfnamefont {J.}~\bibnamefont
  {Schupp}}, \bibinfo {author} {\bibfnamefont {V.}~\bibnamefont {Krcmarsky}},
  \bibinfo {author} {\bibfnamefont {V.}~\bibnamefont {Krutyanskiy}}, \bibinfo
  {author} {\bibfnamefont {M.}~\bibnamefont {Meraner}}, \bibinfo {author}
  {\bibfnamefont {T.}~\bibnamefont {Northup}},\ and\ \bibinfo {author}
  {\bibfnamefont {B.}~\bibnamefont {Lanyon}},\ }\href
  {https://doi.org/10.1103/PRXQuantum.2.020331} {\bibfield  {journal} {\bibinfo
   {journal} {PRX Quantum}\ }\textbf {\bibinfo {volume} {2}},\ \bibinfo {pages}
  {020331} (\bibinfo {year} {2021})}\BibitemShut {NoStop}%
\bibitem [{\citenamefont {Boozer}\ \emph {et~al.}(2007)\citenamefont {Boozer},
  \citenamefont {Boca}, \citenamefont {Miller}, \citenamefont {Northup},\ and\
  \citenamefont {Kimble}}]{Boozer07a}%
  \BibitemOpen
  \bibfield  {author} {\bibinfo {author} {\bibfnamefont {A.~D.}\ \bibnamefont
  {Boozer}}, \bibinfo {author} {\bibfnamefont {A.}~\bibnamefont {Boca}},
  \bibinfo {author} {\bibfnamefont {R.}~\bibnamefont {Miller}}, \bibinfo
  {author} {\bibfnamefont {T.~E.}\ \bibnamefont {Northup}},\ and\ \bibinfo
  {author} {\bibfnamefont {H.~J.}\ \bibnamefont {Kimble}},\ }\href
  {https://doi.org/10.1103/PhysRevLett.98.193601} {\bibfield  {journal}
  {\bibinfo  {journal} {Phys. Rev. Lett.}\ }\textbf {\bibinfo {volume} {98}},\
  \bibinfo {eid} {193601} (\bibinfo {year} {2007})}\BibitemShut {NoStop}%
\bibitem [{\citenamefont {Stute}\ \emph {et~al.}(2013)\citenamefont {Stute},
  \citenamefont {Casabone}, \citenamefont {Brandst\"{a}tter}, \citenamefont
  {Friebe}, \citenamefont {Northup},\ and\ \citenamefont {Blatt}}]{Stute13}%
  \BibitemOpen
  \bibfield  {author} {\bibinfo {author} {\bibfnamefont {A.}~\bibnamefont
  {Stute}}, \bibinfo {author} {\bibfnamefont {B.}~\bibnamefont {Casabone}},
  \bibinfo {author} {\bibfnamefont {B.}~\bibnamefont {Brandst\"{a}tter}},
  \bibinfo {author} {\bibfnamefont {K.}~\bibnamefont {Friebe}}, \bibinfo
  {author} {\bibfnamefont {T.~E.}\ \bibnamefont {Northup}},\ and\ \bibinfo
  {author} {\bibfnamefont {R.}~\bibnamefont {Blatt}},\ }\href
  {https://doi.org/10.1038/nphoton.2012.358} {\bibfield  {journal} {\bibinfo
  {journal} {Nat. Photonics}\ }\textbf {\bibinfo {volume} {7}},\ \bibinfo
  {pages} {219} (\bibinfo {year} {2013})}\BibitemShut {NoStop}%
\bibitem [{\citenamefont {Reiserer}\ \emph {et~al.}(2014)\citenamefont
  {Reiserer}, \citenamefont {Kalb}, \citenamefont {Rempe},\ and\ \citenamefont
  {Ritter}}]{Reiserer2014quantum}%
  \BibitemOpen
  \bibfield  {author} {\bibinfo {author} {\bibfnamefont {A.}~\bibnamefont
  {Reiserer}}, \bibinfo {author} {\bibfnamefont {N.}~\bibnamefont {Kalb}},
  \bibinfo {author} {\bibfnamefont {G.}~\bibnamefont {Rempe}},\ and\ \bibinfo
  {author} {\bibfnamefont {S.}~\bibnamefont {Ritter}},\ }\href
  {http://dx.doi.org/10.1038/nature13177} {\bibfield  {journal} {\bibinfo
  {journal} {Nature}\ }\textbf {\bibinfo {volume} {508}},\ \bibinfo {pages}
  {237} (\bibinfo {year} {2014})}\BibitemShut {NoStop}%
\bibitem [{\citenamefont {Hacker}\ \emph {et~al.}(2016)\citenamefont {Hacker},
  \citenamefont {Welte}, \citenamefont {Rempe},\ and\ \citenamefont
  {Ritter}}]{Hacker2016}%
  \BibitemOpen
  \bibfield  {author} {\bibinfo {author} {\bibfnamefont {B.}~\bibnamefont
  {Hacker}}, \bibinfo {author} {\bibfnamefont {S.}~\bibnamefont {Welte}},
  \bibinfo {author} {\bibfnamefont {G.}~\bibnamefont {Rempe}},\ and\ \bibinfo
  {author} {\bibfnamefont {S.}~\bibnamefont {Ritter}},\ }\href
  {http://dx.doi.org/10.1038/nature18592} {\bibfield  {journal} {\bibinfo
  {journal} {Nature}\ }\textbf {\bibinfo {volume} {536}},\ \bibinfo {pages}
  {193} (\bibinfo {year} {2016})}\BibitemShut {NoStop}%
\bibitem [{\citenamefont {Berman}(1994)}]{Berman94}%
  \BibitemOpen
  \bibinfo {editor} {\bibfnamefont {P.~R.}\ \bibnamefont {Berman}},\ ed.,\
  \href@noop {} {\emph {\bibinfo {title} {Cavity Quantum Electrodynamics}}}\
  (\bibinfo  {publisher} {Academic Press},\ \bibinfo {address} {San Diego},\
  \bibinfo {year} {1994})\BibitemShut {NoStop}%
\bibitem [{\citenamefont {Kimble}(1998)}]{Kimble98}%
  \BibitemOpen
  \bibfield  {author} {\bibinfo {author} {\bibfnamefont {H.}~\bibnamefont
  {Kimble}},\ }\href@noop {} {\bibfield  {journal} {\bibinfo  {journal} {Phys.
  Scr.}\ }\textbf {\bibinfo {volume} {76}},\ \bibinfo {pages} {127} (\bibinfo
  {year} {1998})}\BibitemShut {NoStop}%
\bibitem [{\citenamefont {Stute}\ \emph
  {et~al.}(2012{\natexlab{b}})\citenamefont {Stute}, \citenamefont {Casabone},
  \citenamefont {Brandst\"atter}, \citenamefont {Habicher}, \citenamefont
  {Schmidt}, \citenamefont {Northup},\ and\ \citenamefont {Blatt}}]{Stute12a}%
  \BibitemOpen
  \bibfield  {author} {\bibinfo {author} {\bibfnamefont {A.}~\bibnamefont
  {Stute}}, \bibinfo {author} {\bibfnamefont {B.}~\bibnamefont {Casabone}},
  \bibinfo {author} {\bibfnamefont {B.}~\bibnamefont {Brandst\"atter}},
  \bibinfo {author} {\bibfnamefont {D.}~\bibnamefont {Habicher}}, \bibinfo
  {author} {\bibfnamefont {P.~O.}\ \bibnamefont {Schmidt}}, \bibinfo {author}
  {\bibfnamefont {T.~E.}\ \bibnamefont {Northup}},\ and\ \bibinfo {author}
  {\bibfnamefont {R.}~\bibnamefont {Blatt}},\ }\href
  {http://arxiv.org/abs/1105.0579} {\bibfield  {journal} {\bibinfo  {journal}
  {Appl. Phys. B}\ }\textbf {\bibinfo {volume} {107}},\ \bibinfo {pages} {1145}
  (\bibinfo {year} {2012}{\natexlab{b}})}\BibitemShut {NoStop}%
\bibitem [{\citenamefont {Maunz}\ \emph {et~al.}(2004)\citenamefont {Maunz},
  \citenamefont {Puppe}, \citenamefont {Schuster}, \citenamefont {Syassen},
  \citenamefont {Pinkse},\ and\ \citenamefont {Rempe}}]{Maunz04}%
  \BibitemOpen
  \bibfield  {author} {\bibinfo {author} {\bibfnamefont {P.}~\bibnamefont
  {Maunz}}, \bibinfo {author} {\bibfnamefont {T.}~\bibnamefont {Puppe}},
  \bibinfo {author} {\bibfnamefont {I.}~\bibnamefont {Schuster}}, \bibinfo
  {author} {\bibfnamefont {N.}~\bibnamefont {Syassen}}, \bibinfo {author}
  {\bibfnamefont {P.~W.~H.}\ \bibnamefont {Pinkse}},\ and\ \bibinfo {author}
  {\bibfnamefont {G.}~\bibnamefont {Rempe}},\ }\href
  {https://doi.org/10.1038/nature02387} {\bibfield  {journal} {\bibinfo
  {journal} {Nature}\ }\textbf {\bibinfo {volume} {428}},\ \bibinfo {pages}
  {50} (\bibinfo {year} {2004})}\BibitemShut {NoStop}%
\bibitem [{\citenamefont {Leibrandt}\ \emph {et~al.}(2009)\citenamefont
  {Leibrandt}, \citenamefont {Labaziewicz}, \citenamefont {Vuleti\'{c}},\ and\
  \citenamefont {Chuang}}]{Leibrandt09}%
  \BibitemOpen
  \bibfield  {author} {\bibinfo {author} {\bibfnamefont {D.~R.}\ \bibnamefont
  {Leibrandt}}, \bibinfo {author} {\bibfnamefont {J.}~\bibnamefont
  {Labaziewicz}}, \bibinfo {author} {\bibfnamefont {V.}~\bibnamefont
  {Vuleti\'{c}}},\ and\ \bibinfo {author} {\bibfnamefont {I.~L.}\ \bibnamefont
  {Chuang}},\ }\href {https://doi.org/10.1103/PhysRevLett.103.103001}
  {\bibfield  {journal} {\bibinfo  {journal} {Phys. Rev. Lett.}\ }\textbf
  {\bibinfo {volume} {103}},\ \bibinfo {pages} {103001} (\bibinfo {year}
  {2009})}\BibitemShut {NoStop}%
\bibitem [{\citenamefont {Kim}\ \emph {et~al.}(2012)\citenamefont {Kim},
  \citenamefont {Lee}, \citenamefont {Kim}, \citenamefont {Seo}, \citenamefont
  {Hong}, \citenamefont {Song},\ and\ \citenamefont {An}}]{Kim2012}%
  \BibitemOpen
  \bibfield  {author} {\bibinfo {author} {\bibfnamefont {J.}~\bibnamefont
  {Kim}}, \bibinfo {author} {\bibfnamefont {M.}~\bibnamefont {Lee}}, \bibinfo
  {author} {\bibfnamefont {D.}~\bibnamefont {Kim}}, \bibinfo {author}
  {\bibfnamefont {W.}~\bibnamefont {Seo}}, \bibinfo {author} {\bibfnamefont
  {H.-G.}\ \bibnamefont {Hong}}, \bibinfo {author} {\bibfnamefont
  {Y.}~\bibnamefont {Song}},\ and\ \bibinfo {author} {\bibfnamefont
  {K.}~\bibnamefont {An}},\ }\href {https://doi.org/10.1364/OL.37.001457}
  {\bibfield  {journal} {\bibinfo  {journal} {Opt. Lett.}\ }\textbf {\bibinfo
  {volume} {37}},\ \bibinfo {pages} {1457} (\bibinfo {year}
  {2012})}\BibitemShut {NoStop}%
\bibitem [{\citenamefont {Choi}\ \emph {et~al.}(2010)\citenamefont {Choi},
  \citenamefont {Kang}, \citenamefont {Lim}, \citenamefont {Kim}, \citenamefont
  {Kim}, \citenamefont {Lee},\ and\ \citenamefont {An}}]{Choi2010}%
  \BibitemOpen
  \bibfield  {author} {\bibinfo {author} {\bibfnamefont {Y.}~\bibnamefont
  {Choi}}, \bibinfo {author} {\bibfnamefont {S.}~\bibnamefont {Kang}}, \bibinfo
  {author} {\bibfnamefont {S.}~\bibnamefont {Lim}}, \bibinfo {author}
  {\bibfnamefont {W.}~\bibnamefont {Kim}}, \bibinfo {author} {\bibfnamefont
  {J.-R.}\ \bibnamefont {Kim}}, \bibinfo {author} {\bibfnamefont {J.-H.}\
  \bibnamefont {Lee}},\ and\ \bibinfo {author} {\bibfnamefont {K.}~\bibnamefont
  {An}},\ }\href {https://doi.org/10.1103/PhysRevLett.104.153601} {\bibfield
  {journal} {\bibinfo  {journal} {Phys. Rev. Lett.}\ }\textbf {\bibinfo
  {volume} {104}},\ \bibinfo {pages} {153601} (\bibinfo {year}
  {2010})}\BibitemShut {NoStop}%
\bibitem [{\citenamefont {Kang}\ \emph {et~al.}(2011)\citenamefont {Kang},
  \citenamefont {Lim}, \citenamefont {Hwang}, \citenamefont {Kim},
  \citenamefont {Kim},\ and\ \citenamefont {An}}]{Kang2011}%
  \BibitemOpen
  \bibfield  {author} {\bibinfo {author} {\bibfnamefont {S.}~\bibnamefont
  {Kang}}, \bibinfo {author} {\bibfnamefont {S.}~\bibnamefont {Lim}}, \bibinfo
  {author} {\bibfnamefont {M.}~\bibnamefont {Hwang}}, \bibinfo {author}
  {\bibfnamefont {W.}~\bibnamefont {Kim}}, \bibinfo {author} {\bibfnamefont
  {J.-R.}\ \bibnamefont {Kim}},\ and\ \bibinfo {author} {\bibfnamefont
  {K.}~\bibnamefont {An}},\ }\href {https://doi.org/10.1364/OE.19.002440}
  {\bibfield  {journal} {\bibinfo  {journal} {Opt. Express}\ }\textbf {\bibinfo
  {volume} {19}},\ \bibinfo {pages} {2440} (\bibinfo {year}
  {2011})}\BibitemShut {NoStop}%
\bibitem [{\citenamefont {Kim}\ \emph {et~al.}(2021)\citenamefont {Kim},
  \citenamefont {Kim}, \citenamefont {Lee}, \citenamefont {Shin}, \citenamefont
  {Kang}, \citenamefont {Kim}, \citenamefont {Choi}, \citenamefont {An},\ and\
  \citenamefont {Lee}}]{Kim2021}%
  \BibitemOpen
  \bibfield  {author} {\bibinfo {author} {\bibfnamefont {J.}~\bibnamefont
  {Kim}}, \bibinfo {author} {\bibfnamefont {K.}~\bibnamefont {Kim}}, \bibinfo
  {author} {\bibfnamefont {D.}~\bibnamefont {Lee}}, \bibinfo {author}
  {\bibfnamefont {Y.}~\bibnamefont {Shin}}, \bibinfo {author} {\bibfnamefont
  {S.}~\bibnamefont {Kang}}, \bibinfo {author} {\bibfnamefont {J.-R.}\
  \bibnamefont {Kim}}, \bibinfo {author} {\bibfnamefont {Y.}~\bibnamefont
  {Choi}}, \bibinfo {author} {\bibfnamefont {K.}~\bibnamefont {An}},\ and\
  \bibinfo {author} {\bibfnamefont {M.}~\bibnamefont {Lee}},\ }\href
  {https://doi.org/10.3390/s21186255} {\bibfield  {journal} {\bibinfo
  {journal} {Sensors}\ }\textbf {\bibinfo {volume} {21}},\ \bibinfo {pages}
  {6255} (\bibinfo {year} {2021})}\BibitemShut {NoStop}%
\bibitem [{\citenamefont {Feng}\ and\ \citenamefont {Winful}(2001)}]{Feng2001}%
  \BibitemOpen
  \bibfield  {author} {\bibinfo {author} {\bibfnamefont {S.}~\bibnamefont
  {Feng}}\ and\ \bibinfo {author} {\bibfnamefont {H.~G.}\ \bibnamefont
  {Winful}},\ }\href {https://doi.org/10.1364/OL.26.000485} {\bibfield
  {journal} {\bibinfo  {journal} {Opt. Lett.}\ }\textbf {\bibinfo {volume}
  {26}},\ \bibinfo {pages} {485} (\bibinfo {year} {2001})}\BibitemShut
  {NoStop}%
\bibitem [{\citenamefont {Siegman}(1986)}]{Siegman86}%
  \BibitemOpen
  \bibfield  {author} {\bibinfo {author} {\bibfnamefont {A.~E.}\ \bibnamefont
  {Siegman}},\ }\href@noop {} {\emph {\bibinfo {title} {Lasers}}}\ (\bibinfo
  {publisher} {University Science Books},\ \bibinfo {address} {Sausalito, CA},\
  \bibinfo {year} {1986})\BibitemShut {NoStop}%
\bibitem [{\citenamefont {Hood}\ \emph {et~al.}(2001)\citenamefont {Hood},
  \citenamefont {Kimble},\ and\ \citenamefont {Ye}}]{Hood01}%
  \BibitemOpen
  \bibfield  {author} {\bibinfo {author} {\bibfnamefont {C.~J.}\ \bibnamefont
  {Hood}}, \bibinfo {author} {\bibfnamefont {H.~J.}\ \bibnamefont {Kimble}},\
  and\ \bibinfo {author} {\bibfnamefont {J.}~\bibnamefont {Ye}},\ }\href
  {https://doi.org/10.1103/PhysRevA.64.033804} {\bibfield  {journal} {\bibinfo
  {journal} {Phys. Rev. A}\ }\textbf {\bibinfo {volume} {64}},\ \bibinfo
  {pages} {033804} (\bibinfo {year} {2001})}\BibitemShut {NoStop}%
\bibitem [{\citenamefont {Yeh}(2005)}]{Yeh2005}%
  \BibitemOpen
  \bibfield  {author} {\bibinfo {author} {\bibfnamefont {P.}~\bibnamefont
  {Yeh}},\ }\href@noop {} {\emph {\bibinfo {title} {Optical Waves in Layered
  Media}}}\ (\bibinfo  {publisher} {Wiley},\ \bibinfo {year}
  {2005})\BibitemShut {NoStop}%
\bibitem [{\citenamefont {Garcia}\ \emph {et~al.}(2020)\citenamefont {Garcia},
  \citenamefont {Ferri}, \citenamefont {Reichel},\ and\ \citenamefont
  {Long}}]{Garcia2020}%
  \BibitemOpen
  \bibfield  {author} {\bibinfo {author} {\bibfnamefont {S.}~\bibnamefont
  {Garcia}}, \bibinfo {author} {\bibfnamefont {F.}~\bibnamefont {Ferri}},
  \bibinfo {author} {\bibfnamefont {J.}~\bibnamefont {Reichel}},\ and\ \bibinfo
  {author} {\bibfnamefont {R.}~\bibnamefont {Long}},\ }\href
  {https://doi.org/10.1364/OE.392207} {\bibfield  {journal} {\bibinfo
  {journal} {Opt. Express}\ }\textbf {\bibinfo {volume} {28}},\ \bibinfo
  {pages} {15515} (\bibinfo {year} {2020})}\BibitemShut {NoStop}%
\bibitem [{\citenamefont {Huang}\ and\ \citenamefont
  {Lehmann}(2008)}]{Huang2008}%
  \BibitemOpen
  \bibfield  {author} {\bibinfo {author} {\bibfnamefont {H.}~\bibnamefont
  {Huang}}\ and\ \bibinfo {author} {\bibfnamefont {K.~K.}\ \bibnamefont
  {Lehmann}},\ }\href {https://doi.org/10.1364/AO.47.003817} {\bibfield
  {journal} {\bibinfo  {journal} {Appl. Opt.}\ }\textbf {\bibinfo {volume}
  {47}},\ \bibinfo {pages} {3817} (\bibinfo {year} {2008})}\BibitemShut
  {NoStop}%
\bibitem [{\citenamefont {Barrett}\ \emph {et~al.}(2019)\citenamefont
  {Barrett}, \citenamefont {Barter}, \citenamefont {Stuart}, \citenamefont
  {Yuen},\ and\ \citenamefont {Kuhn}}]{Barrett2019}%
  \BibitemOpen
  \bibfield  {author} {\bibinfo {author} {\bibfnamefont {T.~D.}\ \bibnamefont
  {Barrett}}, \bibinfo {author} {\bibfnamefont {O.}~\bibnamefont {Barter}},
  \bibinfo {author} {\bibfnamefont {D.}~\bibnamefont {Stuart}}, \bibinfo
  {author} {\bibfnamefont {B.}~\bibnamefont {Yuen}},\ and\ \bibinfo {author}
  {\bibfnamefont {A.}~\bibnamefont {Kuhn}},\ }\href
  {https://doi.org/10.1103/PhysRevLett.122.083602} {\bibfield  {journal}
  {\bibinfo  {journal} {Phys. Rev. Lett.}\ }\textbf {\bibinfo {volume} {122}},\
  \bibinfo {pages} {083602} (\bibinfo {year} {2019})}\BibitemShut {NoStop}%
\bibitem [{\citenamefont {Morales}\ \emph {et~al.}(2019)\citenamefont
  {Morales}, \citenamefont {Dreon}, \citenamefont {Li}, \citenamefont
  {Baumg\"artner}, \citenamefont {Zupancic}, \citenamefont {Donner},\ and\
  \citenamefont {Esslinger}}]{Morales2019}%
  \BibitemOpen
  \bibfield  {author} {\bibinfo {author} {\bibfnamefont {A.}~\bibnamefont
  {Morales}}, \bibinfo {author} {\bibfnamefont {D.}~\bibnamefont {Dreon}},
  \bibinfo {author} {\bibfnamefont {X.}~\bibnamefont {Li}}, \bibinfo {author}
  {\bibfnamefont {A.}~\bibnamefont {Baumg\"artner}}, \bibinfo {author}
  {\bibfnamefont {P.}~\bibnamefont {Zupancic}}, \bibinfo {author}
  {\bibfnamefont {T.}~\bibnamefont {Donner}},\ and\ \bibinfo {author}
  {\bibfnamefont {T.}~\bibnamefont {Esslinger}},\ }\href
  {https://doi.org/10.1103/PhysRevA.100.013816} {\bibfield  {journal} {\bibinfo
   {journal} {Phys. Rev. A}\ }\textbf {\bibinfo {volume} {100}},\ \bibinfo
  {pages} {013816} (\bibinfo {year} {2019})}\BibitemShut {NoStop}%
\bibitem [{\citenamefont {Dupr\'e}(2015)}]{Dupre2015}%
  \BibitemOpen
  \bibfield  {author} {\bibinfo {author} {\bibfnamefont {P.}~\bibnamefont
  {Dupr\'e}},\ }\href {https://doi.org/10.1103/PhysRevA.92.053817} {\bibfield
  {journal} {\bibinfo  {journal} {Phys. Rev. A}\ }\textbf {\bibinfo {volume}
  {92}},\ \bibinfo {pages} {053817} (\bibinfo {year} {2015})}\BibitemShut
  {NoStop}%
\bibitem [{\citenamefont {Fleisher}\ \emph {et~al.}(2016)\citenamefont
  {Fleisher}, \citenamefont {Long}, \citenamefont {Liu},\ and\ \citenamefont
  {Hodges}}]{Fleisher2016}%
  \BibitemOpen
  \bibfield  {author} {\bibinfo {author} {\bibfnamefont {A.~J.}\ \bibnamefont
  {Fleisher}}, \bibinfo {author} {\bibfnamefont {D.~A.}\ \bibnamefont {Long}},
  \bibinfo {author} {\bibfnamefont {Q.}~\bibnamefont {Liu}},\ and\ \bibinfo
  {author} {\bibfnamefont {J.~T.}\ \bibnamefont {Hodges}},\ }\href
  {https://doi.org/10.1103/PhysRevA.93.013833} {\bibfield  {journal} {\bibinfo
  {journal} {Phys. Rev. A}\ }\textbf {\bibinfo {volume} {93}},\ \bibinfo
  {pages} {013833} (\bibinfo {year} {2016})}\BibitemShut {NoStop}%
\bibitem [{\citenamefont {Birnbaum}\ \emph {et~al.}(2005)\citenamefont
  {Birnbaum}, \citenamefont {Boca}, \citenamefont {Miller}, \citenamefont
  {Boozer}, \citenamefont {Northup},\ and\ \citenamefont
  {Kimble}}]{Birnbaum05a}%
  \BibitemOpen
  \bibfield  {author} {\bibinfo {author} {\bibfnamefont {K.~M.}\ \bibnamefont
  {Birnbaum}}, \bibinfo {author} {\bibfnamefont {A.}~\bibnamefont {Boca}},
  \bibinfo {author} {\bibfnamefont {R.}~\bibnamefont {Miller}}, \bibinfo
  {author} {\bibfnamefont {A.~D.}\ \bibnamefont {Boozer}}, \bibinfo {author}
  {\bibfnamefont {T.~E.}\ \bibnamefont {Northup}},\ and\ \bibinfo {author}
  {\bibfnamefont {H.~J.}\ \bibnamefont {Kimble}},\ }\href@noop {} {\bibfield
  {journal} {\bibinfo  {journal} {Nature}\ }\textbf {\bibinfo {volume} {436}},\
  \bibinfo {pages} {87} (\bibinfo {year} {2005})}\BibitemShut {NoStop}%
\bibitem [{\citenamefont {Krutyanskiy}\ \emph {et~al.}(2019)\citenamefont
  {Krutyanskiy}, \citenamefont {Meraner}, \citenamefont {Schupp}, \citenamefont
  {Krcmarsky}, \citenamefont {Hainzer},\ and\ \citenamefont
  {Lanyon}}]{Krutyanskiy2019}%
  \BibitemOpen
  \bibfield  {author} {\bibinfo {author} {\bibfnamefont {V.}~\bibnamefont
  {Krutyanskiy}}, \bibinfo {author} {\bibfnamefont {M.}~\bibnamefont
  {Meraner}}, \bibinfo {author} {\bibfnamefont {J.}~\bibnamefont {Schupp}},
  \bibinfo {author} {\bibfnamefont {V.}~\bibnamefont {Krcmarsky}}, \bibinfo
  {author} {\bibfnamefont {H.}~\bibnamefont {Hainzer}},\ and\ \bibinfo {author}
  {\bibfnamefont {B.~P.}\ \bibnamefont {Lanyon}},\ }\href@noop {} {\bibfield
  {journal} {\bibinfo  {journal} {Npj Quantum Inf.}\ }\textbf {\bibinfo
  {volume} {5}},\ \bibinfo {pages} {72} (\bibinfo {year} {2019})}\BibitemShut
  {NoStop}%
\bibitem [{\citenamefont {Briles}\ \emph {et~al.}(2010)\citenamefont {Briles},
  \citenamefont {Yost}, \citenamefont {Cing\"{o}z}, \citenamefont {Ye},\ and\
  \citenamefont {Schibli}}]{Briles2010}%
  \BibitemOpen
  \bibfield  {author} {\bibinfo {author} {\bibfnamefont {T.~C.}\ \bibnamefont
  {Briles}}, \bibinfo {author} {\bibfnamefont {D.~C.}\ \bibnamefont {Yost}},
  \bibinfo {author} {\bibfnamefont {A.}~\bibnamefont {Cing\"{o}z}}, \bibinfo
  {author} {\bibfnamefont {J.}~\bibnamefont {Ye}},\ and\ \bibinfo {author}
  {\bibfnamefont {T.~R.}\ \bibnamefont {Schibli}},\ }\href
  {https://doi.org/10.1364/OE.18.009739} {\bibfield  {journal} {\bibinfo
  {journal} {Opt. Express}\ }\textbf {\bibinfo {volume} {18}},\ \bibinfo
  {pages} {9739} (\bibinfo {year} {2010})}\BibitemShut {NoStop}%
\bibitem [{\citenamefont {Janitz}\ \emph {et~al.}(2017)\citenamefont {Janitz},
  \citenamefont {Ruf}, \citenamefont {Fontana}, \citenamefont {Sankey},\ and\
  \citenamefont {Childress}}]{Janitz2017}%
  \BibitemOpen
  \bibfield  {author} {\bibinfo {author} {\bibfnamefont {E.}~\bibnamefont
  {Janitz}}, \bibinfo {author} {\bibfnamefont {M.}~\bibnamefont {Ruf}},
  \bibinfo {author} {\bibfnamefont {Y.}~\bibnamefont {Fontana}}, \bibinfo
  {author} {\bibfnamefont {J.}~\bibnamefont {Sankey}},\ and\ \bibinfo {author}
  {\bibfnamefont {L.}~\bibnamefont {Childress}},\ }\href
  {https://doi.org/10.1364/OE.25.020932} {\bibfield  {journal} {\bibinfo
  {journal} {Opt. Express}\ }\textbf {\bibinfo {volume} {25}},\ \bibinfo
  {pages} {20932} (\bibinfo {year} {2017})}\BibitemShut {NoStop}%
\bibitem [{\citenamefont {Lee}\ \emph {et~al.}(2019{\natexlab{a}})\citenamefont
  {Lee}, \citenamefont {Lee}, \citenamefont {Hong}, \citenamefont
  {Sch\"uppert}, \citenamefont {Kwon}, \citenamefont {Kim}, \citenamefont
  {Colombe}, \citenamefont {Northup}, \citenamefont {Cho},\ and\ \citenamefont
  {Blatt}}]{Lee2019a}%
  \BibitemOpen
  \bibfield  {author} {\bibinfo {author} {\bibfnamefont {M.}~\bibnamefont
  {Lee}}, \bibinfo {author} {\bibfnamefont {M.}~\bibnamefont {Lee}}, \bibinfo
  {author} {\bibfnamefont {S.}~\bibnamefont {Hong}}, \bibinfo {author}
  {\bibfnamefont {K.}~\bibnamefont {Sch\"uppert}}, \bibinfo {author}
  {\bibfnamefont {Y.-D.}\ \bibnamefont {Kwon}}, \bibinfo {author}
  {\bibfnamefont {T.}~\bibnamefont {Kim}}, \bibinfo {author} {\bibfnamefont
  {Y.}~\bibnamefont {Colombe}}, \bibinfo {author} {\bibfnamefont {T.~E.}\
  \bibnamefont {Northup}}, \bibinfo {author} {\bibfnamefont {D.-I.~D.}\
  \bibnamefont {Cho}},\ and\ \bibinfo {author} {\bibfnamefont {R.}~\bibnamefont
  {Blatt}},\ }\href {https://doi.org/10.1103/PhysRevApplied.12.044052}
  {\bibfield  {journal} {\bibinfo  {journal} {Phys. Rev. Applied}\ }\textbf
  {\bibinfo {volume} {12}},\ \bibinfo {pages} {044052} (\bibinfo {year}
  {2019}{\natexlab{a}})}\BibitemShut {NoStop}%
\bibitem [{\citenamefont {Gallego}\ \emph {et~al.}(2016)\citenamefont
  {Gallego}, \citenamefont {Ghosh}, \citenamefont {Alavi}, \citenamefont {Alt},
  \citenamefont {Martinez-Dorantes}, \citenamefont {Meschede},\ and\
  \citenamefont {Ratschbacher}}]{Gallego2016}%
  \BibitemOpen
  \bibfield  {author} {\bibinfo {author} {\bibfnamefont {J.}~\bibnamefont
  {Gallego}}, \bibinfo {author} {\bibfnamefont {S.}~\bibnamefont {Ghosh}},
  \bibinfo {author} {\bibfnamefont {S.~K.}\ \bibnamefont {Alavi}}, \bibinfo
  {author} {\bibfnamefont {W.}~\bibnamefont {Alt}}, \bibinfo {author}
  {\bibfnamefont {M.}~\bibnamefont {Martinez-Dorantes}}, \bibinfo {author}
  {\bibfnamefont {D.}~\bibnamefont {Meschede}},\ and\ \bibinfo {author}
  {\bibfnamefont {L.}~\bibnamefont {Ratschbacher}},\ }\href
  {https://doi.org/10.1007/s00340-015-6281-z} {\bibfield  {journal} {\bibinfo
  {journal} {Appl. Phys. B}\ }\textbf {\bibinfo {volume} {122}},\ \bibinfo
  {pages} {47} (\bibinfo {year} {2016})}\BibitemShut {NoStop}%
\bibitem [{\citenamefont {Steck}(2010)}]{Steck2010}%
  \BibitemOpen
  \bibfield  {author} {\bibinfo {author} {\bibfnamefont {D.~A.}\ \bibnamefont
  {Steck}},\ }\href@noop {} {\bibfield  {journal} {\bibinfo  {journal}
  {available online at http://steck.us/alkalidata (revision 2.1.4, Dec. 23,
  2010)}\ } (\bibinfo {year} {2010})}\BibitemShut {NoStop}%
\bibitem [{\citenamefont {Ghadimi}\ \emph {et~al.}(2020)\citenamefont
  {Ghadimi}, \citenamefont {Bridge}, \citenamefont {Scarabel}, \citenamefont
  {Connell}, \citenamefont {Shimizu}, \citenamefont {Streed},\ and\
  \citenamefont {Lobino}}]{Ghadimi2020}%
  \BibitemOpen
  \bibfield  {author} {\bibinfo {author} {\bibfnamefont {M.}~\bibnamefont
  {Ghadimi}}, \bibinfo {author} {\bibfnamefont {E.~M.}\ \bibnamefont {Bridge}},
  \bibinfo {author} {\bibfnamefont {J.}~\bibnamefont {Scarabel}}, \bibinfo
  {author} {\bibfnamefont {S.}~\bibnamefont {Connell}}, \bibinfo {author}
  {\bibfnamefont {K.}~\bibnamefont {Shimizu}}, \bibinfo {author} {\bibfnamefont
  {E.}~\bibnamefont {Streed}},\ and\ \bibinfo {author} {\bibfnamefont
  {M.}~\bibnamefont {Lobino}},\ }\href {https://doi.org/10.1364/AO.390881}
  {\bibfield  {journal} {\bibinfo  {journal} {Appl. Opt.}\ }\textbf {\bibinfo
  {volume} {59}},\ \bibinfo {pages} {5136} (\bibinfo {year}
  {2020})}\BibitemShut {NoStop}%
\bibitem [{\citenamefont {Brandst\"atter}\ \emph {et~al.}(2013)\citenamefont
  {Brandst\"atter}, \citenamefont {McClung}, \citenamefont {Sch\"uppert},
  \citenamefont {Casabone}, \citenamefont {Friebe}, \citenamefont {Stute},
  \citenamefont {Schmidt}, \citenamefont {Deutsch}, \citenamefont {Reichel},
  \citenamefont {Blatt},\ and\ \citenamefont {Northup}}]{Brandstaetter2013}%
  \BibitemOpen
  \bibfield  {author} {\bibinfo {author} {\bibfnamefont {B.}~\bibnamefont
  {Brandst\"atter}}, \bibinfo {author} {\bibfnamefont {A.}~\bibnamefont
  {McClung}}, \bibinfo {author} {\bibfnamefont {K.}~\bibnamefont
  {Sch\"uppert}}, \bibinfo {author} {\bibfnamefont {B.}~\bibnamefont
  {Casabone}}, \bibinfo {author} {\bibfnamefont {K.}~\bibnamefont {Friebe}},
  \bibinfo {author} {\bibfnamefont {A.}~\bibnamefont {Stute}}, \bibinfo
  {author} {\bibfnamefont {P.~O.}\ \bibnamefont {Schmidt}}, \bibinfo {author}
  {\bibfnamefont {C.}~\bibnamefont {Deutsch}}, \bibinfo {author} {\bibfnamefont
  {J.}~\bibnamefont {Reichel}}, \bibinfo {author} {\bibfnamefont
  {R.}~\bibnamefont {Blatt}},\ and\ \bibinfo {author} {\bibfnamefont {T.~E.}\
  \bibnamefont {Northup}},\ }\href
  {https://doi.org/http://dx.doi.org/10.1063/1.4838696} {\bibfield  {journal}
  {\bibinfo  {journal} {Rev. Sci. Instrum.}\ }\textbf {\bibinfo {volume}
  {84}},\ \bibinfo {eid} {123104} (\bibinfo {year} {2013})}\BibitemShut
  {NoStop}%
\bibitem [{\citenamefont {Ann}\ \emph {et~al.}(2019)\citenamefont {Ann},
  \citenamefont {Song}, \citenamefont {Kim}, \citenamefont {Yang},\ and\
  \citenamefont {An}}]{Ann2019}%
  \BibitemOpen
  \bibfield  {author} {\bibinfo {author} {\bibfnamefont {B.-m.}\ \bibnamefont
  {Ann}}, \bibinfo {author} {\bibfnamefont {Y.}~\bibnamefont {Song}}, \bibinfo
  {author} {\bibfnamefont {J.}~\bibnamefont {Kim}}, \bibinfo {author}
  {\bibfnamefont {D.}~\bibnamefont {Yang}},\ and\ \bibinfo {author}
  {\bibfnamefont {K.}~\bibnamefont {An}},\ }\href
  {https://doi.org/10.1038/s41598-019-53525-3} {\bibfield  {journal} {\bibinfo
  {journal} {Sci. Rep.}\ }\textbf {\bibinfo {volume} {9}},\ \bibinfo {pages}
  {17110} (\bibinfo {year} {2019})}\BibitemShut {NoStop}%
\bibitem [{\citenamefont {Macha}\ \emph {et~al.}(2020)\citenamefont {Macha},
  \citenamefont {Uru\~nuela}, \citenamefont {Alt}, \citenamefont {Ammenwerth},
  \citenamefont {Pandey}, \citenamefont {Pfeifer},\ and\ \citenamefont
  {Meschede}}]{Macha2020}%
  \BibitemOpen
  \bibfield  {author} {\bibinfo {author} {\bibfnamefont {T.}~\bibnamefont
  {Macha}}, \bibinfo {author} {\bibfnamefont {E.}~\bibnamefont {Uru\~nuela}},
  \bibinfo {author} {\bibfnamefont {W.}~\bibnamefont {Alt}}, \bibinfo {author}
  {\bibfnamefont {M.}~\bibnamefont {Ammenwerth}}, \bibinfo {author}
  {\bibfnamefont {D.}~\bibnamefont {Pandey}}, \bibinfo {author} {\bibfnamefont
  {H.}~\bibnamefont {Pfeifer}},\ and\ \bibinfo {author} {\bibfnamefont
  {D.}~\bibnamefont {Meschede}},\ }\href
  {https://doi.org/10.1103/PhysRevA.101.053406} {\bibfield  {journal} {\bibinfo
   {journal} {Phys. Rev. A}\ }\textbf {\bibinfo {volume} {101}},\ \bibinfo
  {pages} {053406} (\bibinfo {year} {2020})}\BibitemShut {NoStop}%
\bibitem [{\citenamefont {Takahashi}\ \emph {et~al.}(2020)\citenamefont
  {Takahashi}, \citenamefont {Kassa}, \citenamefont {Christoforou},\ and\
  \citenamefont {Keller}}]{Takahashi2020}%
  \BibitemOpen
  \bibfield  {author} {\bibinfo {author} {\bibfnamefont {H.}~\bibnamefont
  {Takahashi}}, \bibinfo {author} {\bibfnamefont {E.}~\bibnamefont {Kassa}},
  \bibinfo {author} {\bibfnamefont {C.}~\bibnamefont {Christoforou}},\ and\
  \bibinfo {author} {\bibfnamefont {M.}~\bibnamefont {Keller}},\ }\href
  {https://doi.org/10.1103/PhysRevLett.124.013602} {\bibfield  {journal}
  {\bibinfo  {journal} {Phys. Rev. Lett.}\ }\textbf {\bibinfo {volume} {124}},\
  \bibinfo {pages} {013602} (\bibinfo {year} {2020})}\BibitemShut {NoStop}%
\bibitem [{\citenamefont {Hunger}\ \emph {et~al.}(2010)\citenamefont {Hunger},
  \citenamefont {Steinmetz}, \citenamefont {Colombe}, \citenamefont {Deutsch},
  \citenamefont {H\"{a}nsch},\ and\ \citenamefont {Reichel}}]{Hunger2010}%
  \BibitemOpen
  \bibfield  {author} {\bibinfo {author} {\bibfnamefont {D.}~\bibnamefont
  {Hunger}}, \bibinfo {author} {\bibfnamefont {T.}~\bibnamefont {Steinmetz}},
  \bibinfo {author} {\bibfnamefont {Y.}~\bibnamefont {Colombe}}, \bibinfo
  {author} {\bibfnamefont {C.}~\bibnamefont {Deutsch}}, \bibinfo {author}
  {\bibfnamefont {T.~W.}\ \bibnamefont {H\"{a}nsch}},\ and\ \bibinfo {author}
  {\bibfnamefont {J.}~\bibnamefont {Reichel}},\ }\href
  {https://doi.org/10.1088/1367-2630/12/6/065038} {\bibfield  {journal}
  {\bibinfo  {journal} {New J. Phys.}\ }\textbf {\bibinfo {volume} {12}},\
  \bibinfo {pages} {065038} (\bibinfo {year} {2010})}\BibitemShut {NoStop}%
\bibitem [{\citenamefont {Neuzner}(2016)}]{Neuzner2016thesis}%
  \BibitemOpen
  \bibfield  {author} {\bibinfo {author} {\bibfnamefont {A.}~\bibnamefont
  {Neuzner}},\ }\emph {\bibinfo {title} {Resonance Fluorescence of an Atom Pair
  in an Optical Resonator}},\ \href@noop {} {Ph.D. thesis},\ \bibinfo  {school}
  {Technische Universit\"at M\"unchen} (\bibinfo {year} {2016})\BibitemShut
  {NoStop}%
\bibitem [{\citenamefont {Ong}\ \emph {et~al.}(2020)\citenamefont {Ong},
  \citenamefont {Sch\"uppert}, \citenamefont {Jobez}, \citenamefont {Teller},
  \citenamefont {Ames}, \citenamefont {Fioretto}, \citenamefont {Friebe},
  \citenamefont {Lee}, \citenamefont {Colombe}, \citenamefont {Blatt},\ and\
  \citenamefont {Northup}}]{Ong2020}%
  \BibitemOpen
  \bibfield  {author} {\bibinfo {author} {\bibfnamefont {F.~R.}\ \bibnamefont
  {Ong}}, \bibinfo {author} {\bibfnamefont {K.}~\bibnamefont {Sch\"uppert}},
  \bibinfo {author} {\bibfnamefont {P.}~\bibnamefont {Jobez}}, \bibinfo
  {author} {\bibfnamefont {M.}~\bibnamefont {Teller}}, \bibinfo {author}
  {\bibfnamefont {B.}~\bibnamefont {Ames}}, \bibinfo {author} {\bibfnamefont
  {D.~A.}\ \bibnamefont {Fioretto}}, \bibinfo {author} {\bibfnamefont
  {K.}~\bibnamefont {Friebe}}, \bibinfo {author} {\bibfnamefont
  {M.}~\bibnamefont {Lee}}, \bibinfo {author} {\bibfnamefont {Y.}~\bibnamefont
  {Colombe}}, \bibinfo {author} {\bibfnamefont {R.}~\bibnamefont {Blatt}},\
  and\ \bibinfo {author} {\bibfnamefont {T.~E.}\ \bibnamefont {Northup}},\
  }\href@noop {} {\bibfield  {journal} {\bibinfo  {journal} {New J. Phys.}\
  }\textbf {\bibinfo {volume} {22}},\ \bibinfo {pages} {063018} (\bibinfo
  {year} {2020})}\BibitemShut {NoStop}%
\bibitem [{\citenamefont {Lee}\ \emph {et~al.}(2019{\natexlab{b}})\citenamefont
  {Lee}, \citenamefont {Friebe}, \citenamefont {Fioretto}, \citenamefont
  {Sch\"uppert}, \citenamefont {Ong}, \citenamefont {Plankensteiner},
  \citenamefont {Torggler}, \citenamefont {Ritsch}, \citenamefont {Blatt},\
  and\ \citenamefont {Northup}}]{Lee2019}%
  \BibitemOpen
  \bibfield  {author} {\bibinfo {author} {\bibfnamefont {M.}~\bibnamefont
  {Lee}}, \bibinfo {author} {\bibfnamefont {K.}~\bibnamefont {Friebe}},
  \bibinfo {author} {\bibfnamefont {D.~A.}\ \bibnamefont {Fioretto}}, \bibinfo
  {author} {\bibfnamefont {K.}~\bibnamefont {Sch\"uppert}}, \bibinfo {author}
  {\bibfnamefont {F.~R.}\ \bibnamefont {Ong}}, \bibinfo {author} {\bibfnamefont
  {D.}~\bibnamefont {Plankensteiner}}, \bibinfo {author} {\bibfnamefont
  {V.}~\bibnamefont {Torggler}}, \bibinfo {author} {\bibfnamefont
  {H.}~\bibnamefont {Ritsch}}, \bibinfo {author} {\bibfnamefont
  {R.}~\bibnamefont {Blatt}},\ and\ \bibinfo {author} {\bibfnamefont {T.~E.}\
  \bibnamefont {Northup}},\ }\href
  {https://doi.org/10.1103/PhysRevLett.122.153603} {\bibfield  {journal}
  {\bibinfo  {journal} {Phys. Rev. Lett.}\ }\textbf {\bibinfo {volume} {122}},\
  \bibinfo {pages} {153603} (\bibinfo {year} {2019}{\natexlab{b}})}\BibitemShut
  {NoStop}%
\bibitem [{\citenamefont {Norcia}\ \emph {et~al.}(2016)\citenamefont {Norcia},
  \citenamefont {Winchester}, \citenamefont {Cline},\ and\ \citenamefont
  {Thompson}}]{Norcia2016}%
  \BibitemOpen
  \bibfield  {author} {\bibinfo {author} {\bibfnamefont {M.~A.}\ \bibnamefont
  {Norcia}}, \bibinfo {author} {\bibfnamefont {M.~N.}\ \bibnamefont
  {Winchester}}, \bibinfo {author} {\bibfnamefont {J.~R.~K.}\ \bibnamefont
  {Cline}},\ and\ \bibinfo {author} {\bibfnamefont {J.~K.}\ \bibnamefont
  {Thompson}},\ }\href@noop {} {\bibfield  {journal} {\bibinfo  {journal} {Sci.
  Adv.}\ }\textbf {\bibinfo {volume} {2}} (\bibinfo {year} {2016})}\BibitemShut
  {NoStop}%
\bibitem [{\citenamefont {Baumann}\ \emph {et~al.}(2010)\citenamefont
  {Baumann}, \citenamefont {Guerlin}, \citenamefont {Brennecke},\ and\
  \citenamefont {Esslinger}}]{Baumann10}%
  \BibitemOpen
  \bibfield  {author} {\bibinfo {author} {\bibfnamefont {K.}~\bibnamefont
  {Baumann}}, \bibinfo {author} {\bibfnamefont {C.}~\bibnamefont {Guerlin}},
  \bibinfo {author} {\bibfnamefont {F.}~\bibnamefont {Brennecke}},\ and\
  \bibinfo {author} {\bibfnamefont {T.}~\bibnamefont {Esslinger}},\ }\href@noop
  {} {\bibfield  {journal} {\bibinfo  {journal} {Nature}\ }\textbf {\bibinfo
  {volume} {464}},\ \bibinfo {pages} {1301} (\bibinfo {year}
  {2010})}\BibitemShut {NoStop}%
\bibitem [{\citenamefont {L\'eonard}\ \emph {et~al.}(2017)\citenamefont
  {L\'eonard}, \citenamefont {Morales}, \citenamefont {Zupancic}, \citenamefont
  {Esslinger},\ and\ \citenamefont {Donner}}]{Leonard2017}%
  \BibitemOpen
  \bibfield  {author} {\bibinfo {author} {\bibfnamefont {J.}~\bibnamefont
  {L\'eonard}}, \bibinfo {author} {\bibfnamefont {A.}~\bibnamefont {Morales}},
  \bibinfo {author} {\bibfnamefont {P.}~\bibnamefont {Zupancic}}, \bibinfo
  {author} {\bibfnamefont {T.}~\bibnamefont {Esslinger}},\ and\ \bibinfo
  {author} {\bibfnamefont {T.}~\bibnamefont {Donner}},\ }\href
  {https://doi.org/10.1038/nature21067} {\bibfield  {journal} {\bibinfo
  {journal} {Nature}\ }\textbf {\bibinfo {volume} {543}},\ \bibinfo {pages}
  {87} (\bibinfo {year} {2017})}\BibitemShut {NoStop}%
\bibitem [{\citenamefont {Hosseini}\ \emph {et~al.}(2017)\citenamefont
  {Hosseini}, \citenamefont {Duan}, \citenamefont {Beck}, \citenamefont
  {Chen},\ and\ \citenamefont {Vuleti\ifmmode~\acute{c}\else
  \'{c}\fi{}}}]{Hosseini2017}%
  \BibitemOpen
  \bibfield  {author} {\bibinfo {author} {\bibfnamefont {M.}~\bibnamefont
  {Hosseini}}, \bibinfo {author} {\bibfnamefont {Y.}~\bibnamefont {Duan}},
  \bibinfo {author} {\bibfnamefont {K.~M.}\ \bibnamefont {Beck}}, \bibinfo
  {author} {\bibfnamefont {Y.-T.}\ \bibnamefont {Chen}},\ and\ \bibinfo
  {author} {\bibfnamefont {V.}~\bibnamefont {Vuleti\ifmmode~\acute{c}\else
  \'{c}\fi{}}},\ }\href {https://doi.org/10.1103/PhysRevLett.118.183601}
  {\bibfield  {journal} {\bibinfo  {journal} {Phys. Rev. Lett.}\ }\textbf
  {\bibinfo {volume} {118}},\ \bibinfo {pages} {183601} (\bibinfo {year}
  {2017})}\BibitemShut {NoStop}%
\bibitem [{\citenamefont {Lee}\ \emph {et~al.}(2022)\citenamefont {Lee},
  \citenamefont {Kim}, \citenamefont {Hong}, \citenamefont {Ha}, \citenamefont
  {Kim}, \citenamefont {Kang}, \citenamefont {Choi}, \citenamefont {An},\ and\
  \citenamefont {Lee}}]{dowon_lee_2022_6324425}%
  \BibitemOpen
  \bibfield  {author} {\bibinfo {author} {\bibfnamefont {D.}~\bibnamefont
  {Lee}}, \bibinfo {author} {\bibfnamefont {M.}~\bibnamefont {Kim}}, \bibinfo
  {author} {\bibfnamefont {J.}~\bibnamefont {Hong}}, \bibinfo {author}
  {\bibfnamefont {T.}~\bibnamefont {Ha}}, \bibinfo {author} {\bibfnamefont
  {J.}~\bibnamefont {Kim}}, \bibinfo {author} {\bibfnamefont {S.}~\bibnamefont
  {Kang}}, \bibinfo {author} {\bibfnamefont {Y.}~\bibnamefont {Choi}}, \bibinfo
  {author} {\bibfnamefont {K.}~\bibnamefont {An}},\ and\ \bibinfo {author}
  {\bibfnamefont {M.}~\bibnamefont {Lee}},\ }\bibfield  {journal} {\bibinfo
  {journal} {https://doi.org/10.5281/zenodo.6324425}\ }\href
  {https://doi.org/10.5281/zenodo.6324425} {10.5281/zenodo.6324425} (\bibinfo
  {year} {2022})\BibitemShut {NoStop}%
\end{thebibliography}%

\clearpage

\appendix*

\section{phase shift at the mirror coating}

\begin{figure*}[htbp]
	\centering
	\centering\includegraphics[width=13.0cm]{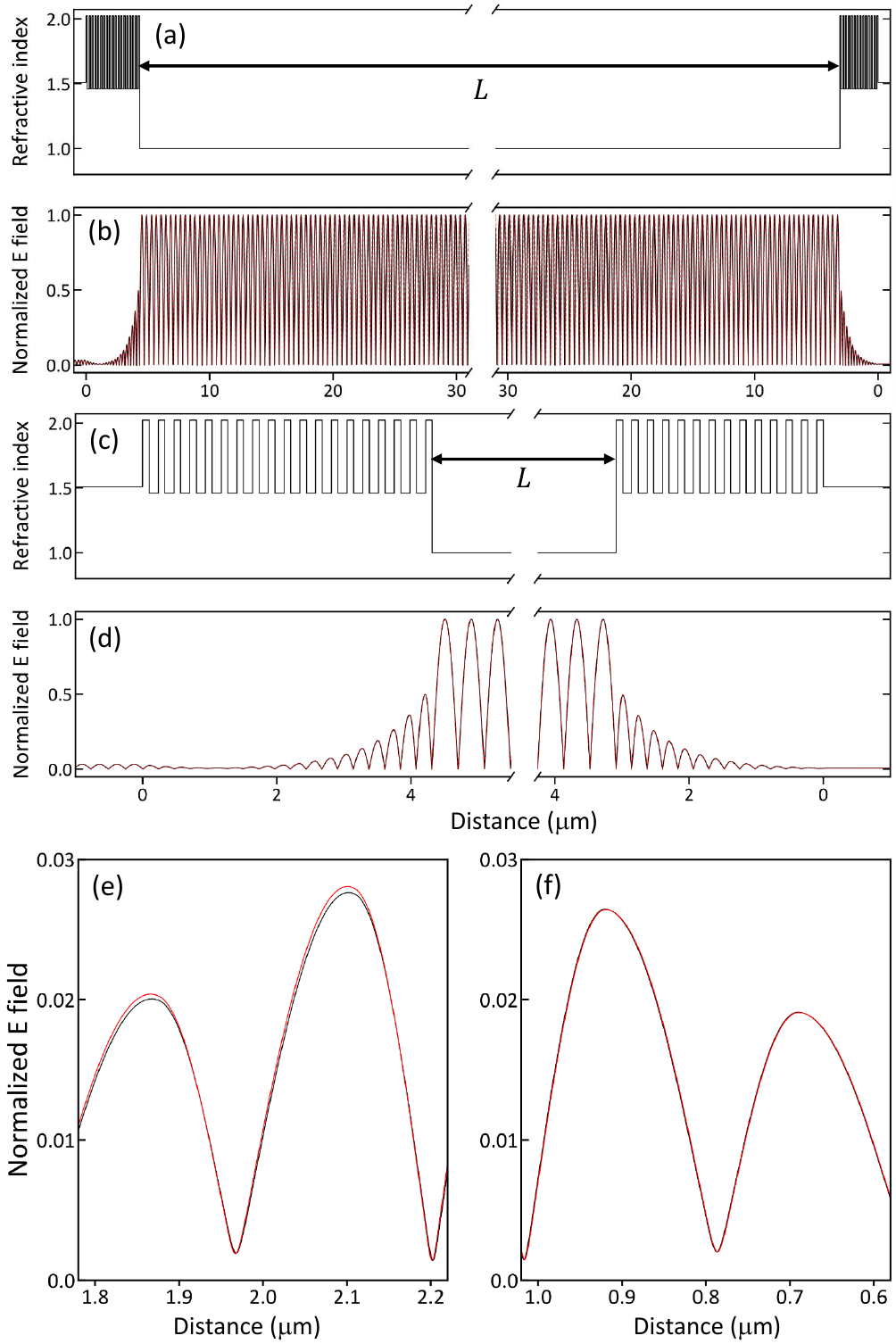} 
	\caption{(a), (c) Refractive index of SiO$_{2}$ and Ta$_{2}$O$_{5}$ coating layers and fused silica substrate. 
	Low transmission mirror is at left and high transmission loss mirror at right. 
	(b), (d) Normalized electric field in the cavity. The field at 782~nm is shown in black and at 780~nm in red. 
	Zoomed-in view of electric fields in the mirror coatings in the low transmission mirror (e) and high transmission mirror (f). 
	The difference of the field between 782 and 780~nm is more pronounced in the low transmission mirror.}
	\label{sfig:TMM}
\end{figure*}

We estimate the phase difference $\Delta\phi_{\rm{coat}} = \phi_{(n+1), 00} - \phi_{n, 00}$ occurred at the mirror coating.
The phase shift at 780~nm is denoted by $\phi_{(n+1), 00}$ and $\phi_{n, 00}$ at 782~nm.
Figs.~\ref{sfig:TMM}(a) and (c) show the coating layers of two mirrors with an actual mirror spacing $L$.
The mirror with low transmission loss (left) consists of total 37 dielectric layers, which are alternating with Ta$_{2}$O$_{5}$ and SiO$_{2}$ films of each thickness $\lambda_{\rm{1}}/n_{\rm{H}}$ and $\lambda_{\rm{1}}/n_{\rm{L}}$, deposited on a fused silica substrate.
The central coating wavelength $\lambda_{1}$ is 791~nm, refractive index of Ta$_{2}$O$_{5}$ is $n_{\rm{H}}=2.0411$, and that of SiO$_{2}$ is $n_{\rm{L}}=1.4550$. 
The fused silica has $n_{\rm{sub}} = 1.5098$.
All of these values are adapted from Ref.~\cite{Hood01} with the indication that they could vary by $\sim1$\%.
It should also be taken into account that the central coating wavelength has a similar level of uncertainty. 
The mirror with high transmission loss (right) is made of 27 alternating layers of Ta$_{2}$O$_{5}$ and SiO$_{2}$, with a coating wavelength of $\lambda_{2} = 786$~nm.
In both cases the top layer of our mirrors is Ta$_{2}$O$_{5}$.

We carry out a numerical simulation to estimate the electric field in the cavity. 
The field is calculated with the scheme of transfer matrix method (TMM)~\cite{Yeh2005}, in which the transmittance (T) and reflectance (R) of all layers are multiplied to estimate the total T/R of the mirror and cavity.
The aim of the simulation is \textit{to find a set of parameters $\lambda_{1}, \lambda_{2}, n_{\rm{H}}$, $n_{\rm{L}}$, and $L$} in such a way that \textit{the calculated frequencies agree with the two measured frequencies,} $\nu_{1} = \nu_{n,00} - \nu_{\rm{trans}} = 383.22230$~THz and $\nu_{2} = \nu_{(n+1),00} - \nu_{\rm{trans}} = 384.21050$~THz.
Since the effect of radius of curvature (ROC) is not included in the TMM simulation, we omit the contribution stemmed from ROC ($\nu_{\rm{trans}}=17.27$~GHz) in this estimation.
In principle, it is enough to scan two parameters, one of the coating parameters ($\lambda_{1}, \lambda_{2}, n_{\rm{H}}$, $n_{\rm{L}}$) and $L$ to obtain $\nu_{1}$ and $\nu_{2}$: However, since we aim to find the coating parameters close to more realistic condition, we scan all coating parameters over the ranges of $\sim1$\%. 
This numerical calculation finds the coating parameters and $L$ that result in $\nu_{1}$ and $\nu_{2}$.
Then, the frequency difference $\nu_{2}-\nu_{1}$ gives the coating phase difference $\Delta\phi_{\rm{coat}}$ between 780 and 782~nm (refer Eq.~(1) of main text), i.e., $\Delta\phi_{\rm{coat}}  = 2\pi \cdot (1 - \nu_{\rm{long}} / \nu_{\rm{fsr}})$, where $\nu_{\rm{long}} = \nu_{2}-\nu_{1}$.

The simulation is performed in the following way: Decreasing $L$ from $L_{\rm{eff}} = 151.686$~$\mu$m (upper limit of our cavity length), we scan the four coating parameters over $\pm1$\% with respect to the given central values. 
When $L=150.978$~$\mu$m, we find that the two measured frequencies $\nu_{1}$ and $\nu_{2}$ are accurately obtained with $n_{\rm{H}}=2.0227$, $n_{\rm{L}}=1.4608$, $\lambda_{1}=795.5$~nm, and $\lambda_{2}=780.1$~nm, which gives $\Delta\phi_{\rm{coat}} = 29.3132$~mrad.

\end{document}